\documentclass[useAMS,usenatbib]{mn2e}

\usepackage{epsfig,color,pifont}
\usepackage{natbib}
\bibpunct{(}{)}{;}{a}{}{,}
\def\spose#1{\hbox to 0pt{#1\hss}}
\def\ltsimm{\mathrel{\spose{\lower 3pt\hbox{$\sim$}}
        \raise 2.0pt\hbox{$<$}}}
\def\gtsimm{\mathrel{\spose{\lower 3pt\hbox{$\sim$}}
        \raise 2.0pt\hbox{$>$}}}
\def\Mdot{\hbox{${\dot M}$}}

\def\km{{\rm\thinspace km}}
\def\cm{{\rm\thinspace cm}}

\def\s{{\rm\thinspace s}}
\def\yr{{\rm\thinspace yr}}
\def\g{{\rm\thinspace g}}
\def\kmps{\hbox{${\rm\km\s^{-1}\,}$}}

\def\erg{{\rm\thinspace erg}}

\def\Hz{{\rm\thinspace Hz}}

\def\ster{{\rm\thinspace ster}}
\def\ergps{\hbox{${\rm\erg\s^{-1}\,}$}}

\def\Msol{\hbox{${\rm\thinspace M_{\odot}}$}}

\def\Msolpyr{\hbox{${\rm\Msol\yr^{-1}\,}$}}
\def\pcm{\hbox{${\rm\cm^{-1}\,}$}}
\def\pcm2{\hbox{${\rm\cm^{-2}\,}$}}
\def\pcm3{\hbox{${\rm\cm^{-3}\,}$}}
\def\ergpscm3Hz{\hbox{${\rm\ergps\cm^{-3}\Hz^{-1}\,}$}}
\def\ergpscm3Hzster{\hbox{${\rm\ergps\cm^{-3}\Hz^{-1}\ster^{-1}\,}$}}
\def\gpcm3{\hbox{${\rm\g\cm^{-3}\,}$}}
\def\ergpcm2{\hbox{${\rm\erg\cm^{-2}\,}$}}
\def\ergpcm3{\hbox{${\rm\erg\cm^{-3}\,}$}}
\def\phpscm2{\hbox{${\rm photons\s^{-1}\cm^{-2}\,}$}}

\newcommand{\tick}{\ding{52}}  %from the pifont package
\title[Turbulent Cloud Destruction]{The Turbulent Destruction of Clouds - II. Mach Number Dependence, Mass-loss Rates, and Tail Formation}
\author[J.~M.~Pittard, T.~W.~Hartquist, S.~A.~E.~G.~Falle]
{J. M. Pittard$^{1}$\thanks{E-mail: jmp@ast.leeds.ac.uk}, T. W. Hartquist$^{1}$, and S. A. E. G. Falle$^{2}$\\
$^{1}$School of Physics and Astronomy, The University of
        Leeds, Leeds LS2 9JT, UK\\
$^{2}$Department of Applied Mathematics, The University of
        Leeds, Leeds LS2 9JT, UK\\
}

\begin{document}

\date{Accepted 2010 February 10. Received 2010 February 10; in original form 2009 September 22}

\pagerange{\pageref{firstpage}--\pageref{lastpage}} \pubyear{2010}

\maketitle

\label{firstpage}

\begin{abstract}
The turbulent destruction of a cloud subject to the passage of an
adiabatic shock is studied. We find large discrepancies between the
lifetime of the cloud and the analytical result of
\citet{Hartquist:1986}. These differences appear to be due to the
assumption in \citet{Hartquist:1986} that mass-loss occurs largely as
a result of lower pressure regions on the surface of the cloud away
from the stagnation point, whereas in reality Kelvin-Helmholtz (KH)
instabilities play a dominant role in the cloud destruction. We find
that the true lifetime of the cloud (defined as when all of the
material from the core of the cloud is well mixed with the intercloud
material in the hydrodynamic cells) is about $6 \times t_{\rm KHD}$,
where $t_{\rm KHD}$ is the growth timescale for the most disruptive,
long-wavelength, KH instabilities.  These findings have wide
implications for diffuse sources where there is transfer of material
between hot and cool phases.

The properties of the interaction as a function of Mach number and
cloud density contrast are also studied. The interaction is milder at
lower Mach numbers with the most marked differences occuring at low
shock Mach numbers when the postshock gas is subsonic with respect to
the cloud (i.e. $M < 2.76$).  Material stripped off the cloud only
forms a long ``tail-like'' feature if $\chi \gtsimm 10^{3}$.
\end{abstract}

\begin{keywords}
hydrodynamics -- ISM: clouds -- ISM: kinematics and dynamics -- shock waves -- supernova remnants -- turbulence
\end{keywords}

\section{Introduction}
\label{sec:intro}
The interchange of material between dense cool phases and a hotter,
more tenuous, external medium is a key phenomenon in astrophysics
which influences the morphology and evolution of diffuse sources over
a wide range of spatial and energy scales. Clouds (also referred to as
clumps, globules, and knots) may either accumulate material from, or
lose material to, the surrounding medium. In the latter case,
mass-loss can occur through hydrodynamic ablation, or thermal or
photoionized evaporation \citep[see, e.g.,][and references
therein]{Pittard:2007}. The response of such multi-phase environments
to the impact of winds and shocks is central to the investigation of
feedback in star and galaxy formation, and encompasses objects such
as tails behind mass-losing stars
\citep[e.g.][]{Martin:2007}, planetary nebulae
\citep[e.g.][]{Matsuura:2009}, supernova remnants
\citep*[e.g.][]{Danforth:2001,Levenson:2002}, starburst regions
\citep[e.g.][]{Westmoquette:2007,Westmoquette:2009}, starburst
superwinds \citep*[e.g.][]{Strickland:2000b,Cecil:2002,Veilleux:2005},
and the intracluster medium \citep*[e.g.][]{Concelise:2001}.

Such interactions have been extensively investigated using numerical
hydrodynamics over the last two decades. A large literature of
shock-cloud, wind-cloud, and jet-cloud interactions now exists. In it,
the effects of radiative cooling, thermal conduction, and ordered
magnetic fields have all been considered. These studies have provided
great insight into the behaviour and evolution of clouds subject to a
variety of internal and external conditions. However, all suffer from
a common and fundamental problem. In most astrophysical environments
the interaction of shocks, winds, and jets with clouds takes place at
high Reynolds numbers and is turbulent \citep[see][for a review of the
key physics]{Pittard:2009}.  Unfortunately, the formation of fully
developed turbulence is prevented by the artificial viscosity inherent
in the hydrodynamical simulations. The turbulent mixing of cloud
material into the surrounding medium is therefore limited, and the
destruction of the cloud is hindered.

The only tractable solution to this problem is to use a statistical
approach to describe the turbulent flow. One possible method involves the use
of a subgrid turbulent viscosity model, designed to calculate the
properties of the turbulence and the resulting increase in the
transport coefficients that the turbulence brings. In this way, the
subgrid turbulence model emulates a high Reynolds number flow.  An
initial investigation by \citet{Pittard:2009} of an adiabatic, Mach 10,
shock-cloud interaction found increasing and significant differences
between the results from a $k$-$\epsilon$ turbulence model and an
inviscid model as the density contrast between the cloud and
inter-cloud medium, $\chi$, increased past $10^2$. In addition, the
$k$-$\epsilon$ turbulence model demonstrated better convergence in
resolution tests, a feature which is particularly useful for
simulations involving multiple clouds.  The effect of a turbulent, as
opposed to a laminar, post-shock medium sweeping over the cloud was
also investigated. The strong ``buffeting'' that the
cloud receives in this case results in its significantly quicker
destruction.

It is also surprising that previous shock-cloud studies have focussed
almost exclusively on high Mach number interactions (see
Table~\ref{tab:previous}). Where lower Mach numbers have been
considered \citep[e.g.,][]{Fragile:2004,Nakamura:2006}, a detailed
comparison to higher Mach number interactions is lacking.  In this
work we extend the shock-cloud investigation of \citet{Pittard:2009}
by examining the Mach number dependence of the interaction.  We
compare our numerical results against the analytical mass-loss rate
formula of \citet{Hartquist:1986}, and for the first time investigate
in detail how well it describes the rate of destruction of the
cloud. We also determine the conditions which are required for a long
identifiable tail to be created behind the cloud.

The structure of this paper is as follows. In Section~\ref{sec:setup}
the numerical method and analysis are
introduced. Section~\ref{sec:stages} notes the main stages in the
destruction of a cloud by a high Mach number shock, and the
Mach-scaling of the interaction is discussed. Our results are
presented in Section~\ref{sec:results}, where a variety of simulations
for strong and weak shocks are presented and
compared. Section~\ref{sec:summary} summarizes and concludes this
work.

\begin{table*}
\begin{center}
\caption[]{A representative (but incomplete) summary of previous numerical investigations of shock-cloud
and wind-cloud interactions. $\chi$ is the density contrast of the
cloud with respect to the ambient medium and $M$ is the shock Mach
number. The interaction types are: SCS = shock-cloud (single); SCM = shock-cloud (multiple);
WCS = wind-cloud (single). The references are as follows: 
$^a$\citet{Stone:1992}; $^b$\citet{Klein:1994}; $^c$\citet{MacLow:1994}; $^d$\citet{Mellema:2002}; $^e$\citet{Fragile:2004}; $^f$\citet{Orlando:2005}; $^g$\citet{Nakamura:2006}; $^h$\citet{vanLoo:2007}; $^i$\citet{Orlando:2008}; $^j$\citet{Shin:2008}; $^k$\citet{Poludnenko:2002}; $^l$\citet{Gregori:2000}; $^m$\citet{Poludnenko:2004}; $^n$\citet{Marcolini:2005}; $^o$\citet{Raga:2007}; $^p$\citet{Vieser:2007}.}
\label{tab:previous}
\begin{tabular}{llllllccc}
\hline
\hline
Authors & Interaction & Geometry & Typical (max) & $\chi$ & $M$ & Cooling? & Conduction? & Magnetic \\
 & type & & resolution & & & & & fields?\\
\hline
SN92$^a$ & SCS & 3D XYZ & 60 (60) & 10 & 10 & & & \\
KMC94$^b$ & SCS & 2D RZ & 120 (240) & 3,10,30, & 10,100,1000 & & & \\
 & & & & 100,400 \\
MC94$^c$ & SCS & 2D RZ,XY & 50 (240) & 10 & 10,100 & & & \tick \\ 
MKR02$^d$ & SCS & 2D RZ,XY & 200 (200) & 1000 & 10 & \tick & & \\
F04$^e$ & SCS & 2D XY & 200 (200) & 1000 & 5,10,20,40 & \tick & & \\
O05$^f$ & SCS & 2D RZ, 3D XYZ & 132 (132) & 10 & 30,50 & \tick & \tick & \\
N06$^g$ & SCS & 2D RZ, 3D XYZ & 120 (960) & 10,100 & 1.5,10,100,1000 & & & \\
V07$^h$ & SCS & 2D RZ & 640 (640) & 45 & 1.5,2.5,5 & \tick & & \tick \\
O08$^i$ & SCS & 2.5D XYZ & 132 (528) & 10 & 50 & \tick & \tick & \tick \\
SSS08$^j$ & SCS & 3D XYZ & 68 (68) & 10 & 10 & & & \tick \\
PFB02$^k$ & SCM & 2D XY & 32 (32) & 500 & 10 \\
G00$^l$ & WCS & 3D XYZ & 16 (26) & 10,30,100 & 1.5,3 & & & \tick \\
PFM04$^m$ & WCS & 2D RZ & 128 (128) & 100 & 10-200 & \tick & & \\
M05$^n$ & WCS & 2D RZ & 150 (150) & 100,500 & 3,6.67 & \tick & \tick & \\
R07$^o$ & WCS & 3D XYZ & 77 (77) & 10 & 242 & \tick & & \\
VH07$^p$ & WCS & 2D RZ & 33 (51) & 0.6,1.2, & 0.3 & \tick & \tick & \\
 & & & & $6.1 \times10^{4}$ \\
\hline
\end{tabular}
\end{center}
%\medskip
\flushleft{Notes: \citet{Patnaude:2005} and \citet{Cooper:2009} 
consider a cloud with
substructure.  Multiple-cloud interactions were also considered by
\citet{Fragile:2004} and \citet{Melioli:2005}, though this was not the
main focus of their work.  Simulations with continuous mass-injection
(to mimic the destruction of very long lived clouds) have been
presented by \citet{Falle:2002}, \citet{Pittard:2005},
\citet{Dyson:2006}, and \citet{Pope:2008}. \citet{Pittard:2005}
investigated the interaction of a wind with multiple mass-injection
sources.  \\}
\end{table*}

\begin{figure*}
%paper scales in x.dat: without scales = 3.3x12.0, with scales = 4.5x12.0
\psfig{figure=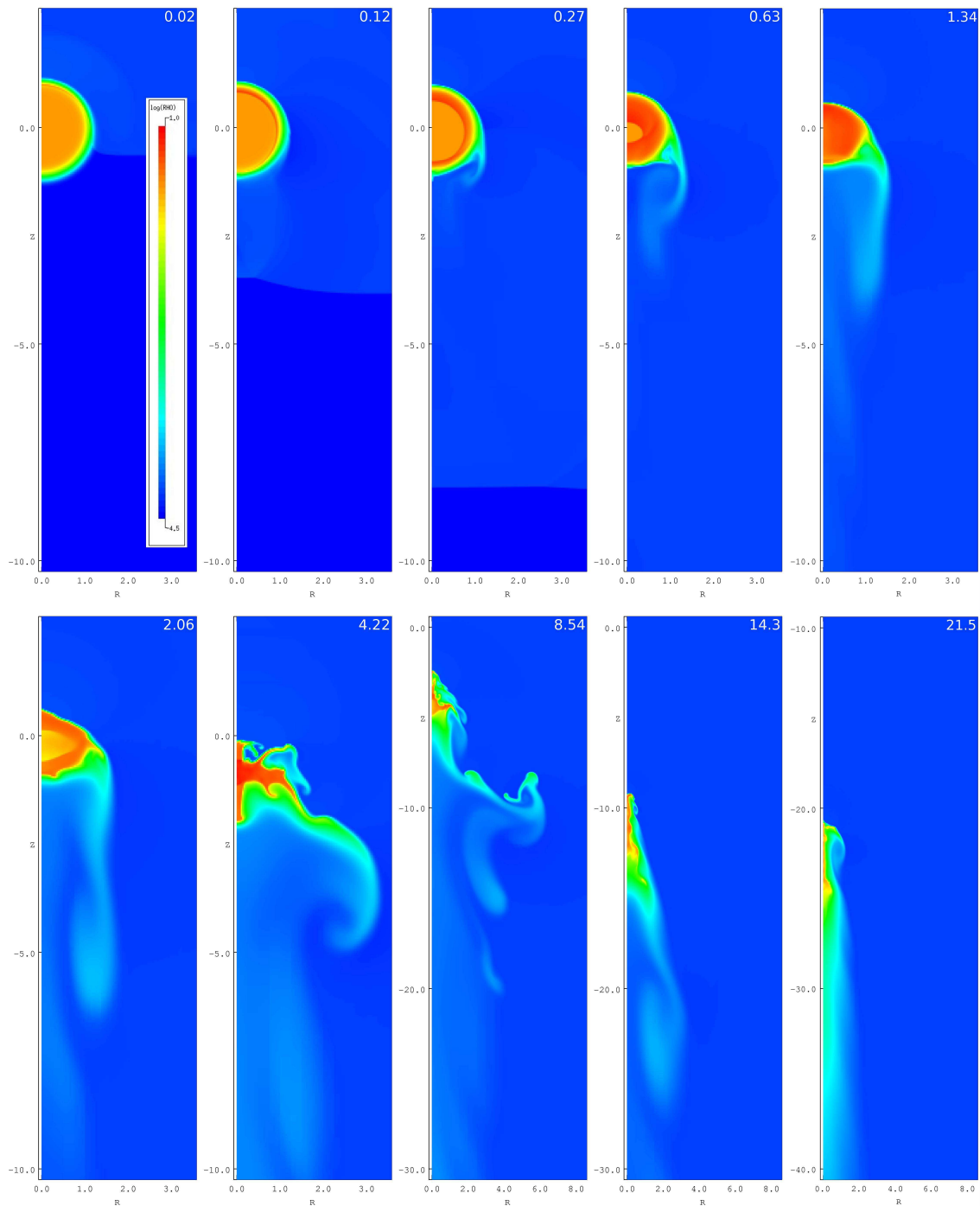,width=15.00cm}
\caption[]{Snapshots of the density distribution from a $k$-$\epsilon$
calculation of a Mach 1.5 adiabatic shock hitting a cloud with a
density contrast of $10^{3}$ with respect to the ambient medium (model
m1.5c3). The evolution proceeds left to right and top to bottom with
$t = 0.02, 0.12, 0.27, 0.63, 1.34, 2.06, 4.22, 8.54, 14.3$, and
$21.5\;t_{\rm cc}$.  Note the different spatial scale used in the last
three panels. The colour scale is ${\rm log_{10}}\,\rho$, and spans the
range $-4.5$ (blue) to $+1.0$ (red).} 
\label{fig:m1.5lokeps}
\end{figure*}

\begin{figure*}
%paper scales in x.dat: without scales = 3.3x12.0, with scales = 4.5x12.0
\psfig{figure=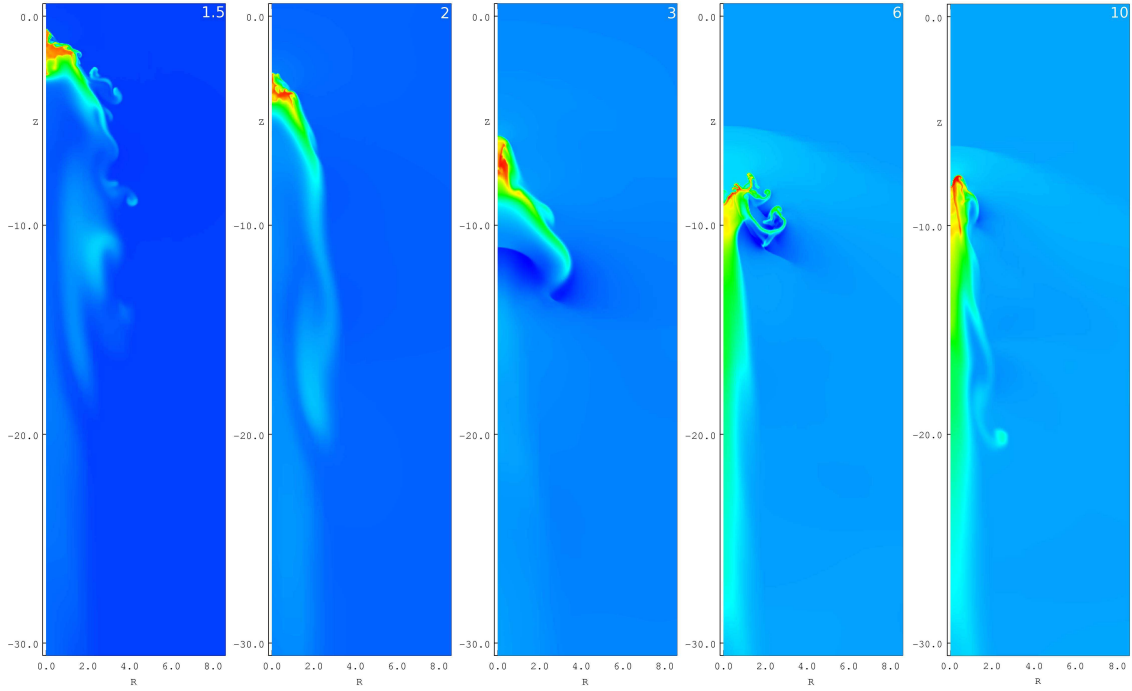,width=15.00cm}
\caption[]{Comparison of the mass density distributions at $t =
  5.66\;t_{\rm cc}$ for $k$-$\epsilon$ subgrid turbulence
  calculations of $\chi=10^{3}$ clouds hit by adiabatic shocks with
  Mach numbers of 1.5 (left), 2, 3, 6, and 10 (right). The density colour scale
is the same as in Fig.~\ref{fig:m1.5lokeps}.}
\label{fig:machcomplo}
\end{figure*}

\begin{figure*}
%paper scales in x.dat: without scales = 3.3x12.0, with scales = 4.5x12.0
\psfig{figure=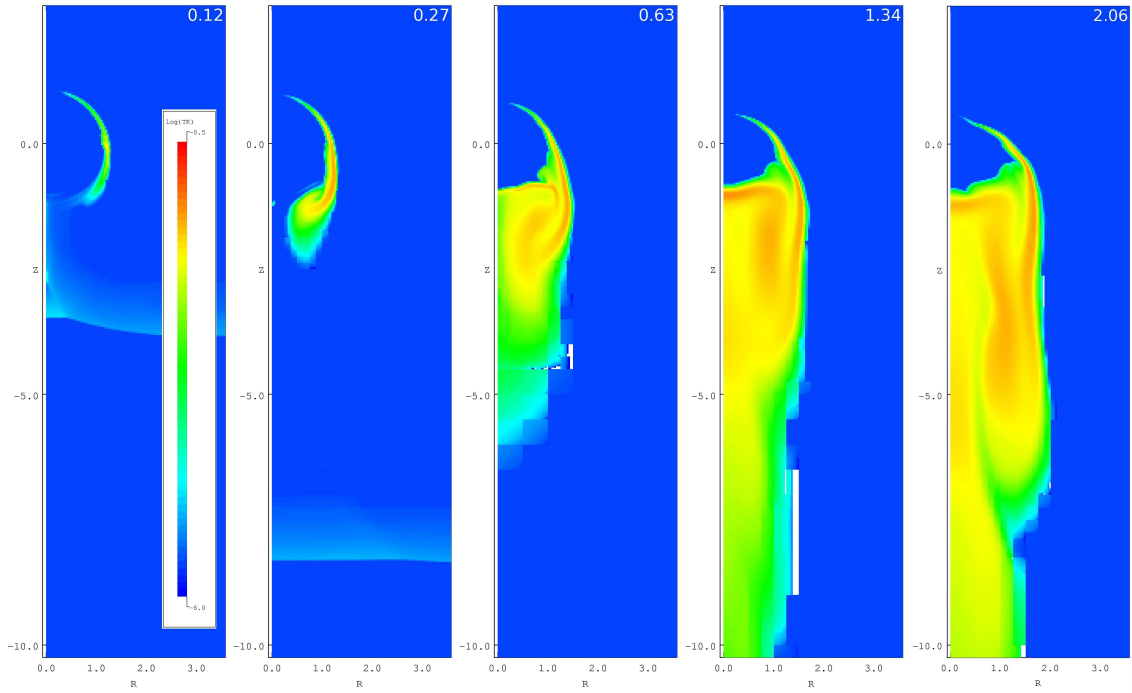,width=15.0cm}
\caption[]{Snapshots of the turbulent energy per unit mass, $k$, from
the $k$-$\epsilon$ calculation with $\chi=10^{3}$ and
$M=1.5$ (model m1.5c3). The evolution proceeds left to right with
$t = 0.12, 0.27, 0.63, 1.34$, and $2.06\;t_{\rm cc}$. The white rectangular regions
are artefacts of the plotting routine.}
\label{fig:m1.5tk}
\end{figure*}

\begin{figure*}
%paper scales in x.dat: without scales = 3.3x12.0, with scales = 4.5x12.0
\psfig{figure=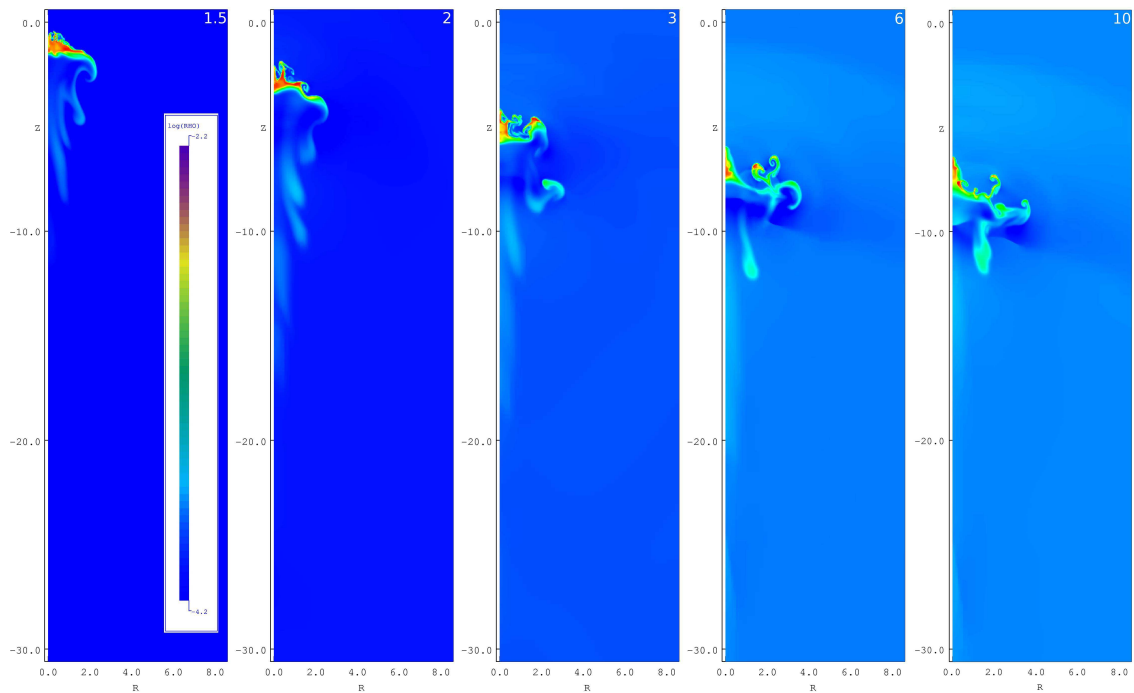,width=15.0cm}
\caption[]{Comparison of the mass density distributions from
$k$-$\epsilon$ subgrid turbulence models at $t = 4.85\;t_{\rm cc}$ for
clouds with $\chi=10^{2}$ hit by adiabatic shocks with Mach numbers
of 1.5 (left), 2, 3, 6, and 10 (right). }
\label{fig:dens1e2machcomplo}
\end{figure*}

\begin{table*}
\begin{center}
\caption[]{The dependences of the global cloud and core properties on
the shock Mach number in subgrid turbulence calculations with cloud
density contrasts of $\chi=10$ (models ``c1''), $\chi=10^{2}$ (models
``c2''), and $\chi=10^{3}$ (models ``c3'').  The Mach number for each
model is given by the number following the initial ``m'' in the model
name - calculations with Mach numbers of 1.5, 2, 3, 4, 6, 10, and 40
were made.  The time-dependent quantities are evaluated at $t=t_{\rm
mix}$ \citep[cf.][]{Pittard:2009}. Values in parentheses are obtained
from integrations over the ``core'' mass rather than the ``cloud''
mass.}
\label{tab:resultslo}
\begin{tabular}{lccccccc}
\hline
\hline
Model & $t_{\rm drag}/t_{\rm cc}$ & $t_{\rm mix}/t_{\rm cc}$ & $a/r_{\rm c}$ & $c/r_{\rm c}$ & $c/a$ & $\langle\rho\rangle/\rho_{\rm max}$ & $\langle v_{\rm z}\rangle/v_{\rm b}$ \\
\hline
m1.5c1 & 2.97 (3.04) & (10.3) & 1.90 (1.97) & 2.76 (2.74) & 1.45 (1.39) & 0.364 (0.543) & 0.292 (0.305) \\
m2c1 & 1.94 (1.98) & (7.95) & 1.71 (1.78) & 2.68 (2.79) & 1.57 (1.57) & 0.452 (0.655) & 0.404 (0.432) \\
m3c1 & 1.40 (1.47) & (6.54) & 1.52 (1.35) & 2.55 (2.75) & 1.68 (2.03) & 0.580 (0.909) & 0.471 (0.472) \\
m4c1 & 1.18 (1.25) & (6.11) & 1.67 (1.58) & 2.46 (2.37) & 1.47 (1.50) & 0.649 (1.008) & 0.541 (0.537) \\
m6c1 & 1.04 (1.13) & (6.10) & 1.77 (1.78) & 2.54 (2.70) & 1.44 (1.51) & 0.641 (0.883) & 0.604 (0.629) \\
m10c1 & 1.04 (1.11) & (5.96) & 1.66 (1.74) & 2.20 (2.40) & 1.33 (1.38) & 0.680 (1.030) & 0.579 (0.629) \\
m40c1 & 0.91 (1.01) & (6.20) & 1.78 (1.89) & 2.26 (2.42) & 1.27 (1.28) & 0.692 (1.044) & 0.599 (0.657) \medskip \\

m1.5c2 & 6.04 (14.0) & (8.41) & 2.68 (1.46) & 9.63 (1.57) & 3.59 (1.07) & 0.046 (0.197) & 0.195 (0.111) \\
m2c2 & 4.43 (4.70) & (6.70) & 3.11 (2.96) & 8.75 (3.35) & 2.81 (1.13) & 0.051 (0.109) & 0.354 (0.307) \\
m3c2 & 3.61 (3.78) & (6.04) & 4.32 (4.64) & 6.90 (1.79) & 1.60 (0.38) & 0.056 (0.129) & 0.511 (0.490) \\
m4c2 & 3.44 (3.49) & (5.55) & 4.09 (4.40) & 5.29 (2.20) & 1.29 (0.50) & 0.063 (0.121) & 0.536 (0.535) \\
m6c2 & 3.27 (3.32) & (4.98) & 3.95 (3.28) & 4.62 (1.98) & 1.17 (0.60) & 0.073 (0.181) & 0.591 (0.517) \\
m10c2 & 3.08 (3.09) & (4.86) & 3.92 (3.35) & 4.12 (3.05) & 1.05 (0.91) & 0.075 (0.142) & 0.593 (0.554) \\
m40c2 & 2.96 (3.01) & (4.63) & 3.88 (3.64) & 4.04 (2.29) & 1.04 (0.63) & 0.077 (0.163) & 0.586 (0.542) \medskip \\

m1.5c3 & 9.19 (29.7) & (13.3) & 5.20 (1.66) & 65.5 (4.20) & 12.6 (2.53) & 0.0038 (0.029) & 0.199 (0.073) \\
m2c3 & 8.66 (13.1) & (10.5) & 3.42 (2.55) & 57.4 (5.86) & 16.8 (2.29) & 0.0051 (0.019) & 0.263 (0.142) \\ 
m3c3 & 7.92 (9.73) & (10.3) & 4.31 (4.52) & 67.5 (9.96) & 15.7 (2.21) & 0.0054 (0.012) & 0.430 (0.360) \\
m4c3 & 6.98 (10.2) & (8.34) & 3.40 (2.30) & 61.2 (6.07) & 18.0 (2.64) & 0.0069 (0.027) & 0.394 (0.213) \\
m6c3 & 6.07 (6.72) & (6.95) & 4.19 (3.78) & 42.1 (5.02) & 10.1 (1.33) & 0.0073 (0.019) & 0.449 (0.333) \\
m10c3 & 6.84 (7.15) & (7.83) & 4.01 (2.82) & 49.1 (6.88) & 12.2 (2.44) & 0.0078 (0.025) & 0.430 (0.271) \\
m40c3 & 5.29 (5.63) & (6.10) & 4.47 (4.29) & 31.1 (5.52) & 6.95 (1.29) & 0.0078 (0.018) & 0.467 (0.366) \\
\hline
\end{tabular}
\end{center}
%\medskip
\end{table*}

\section{The numerical setup}
\label{sec:setup}
The calculations were performed in 2D axisymmetry using an Eulerian
adapative mesh refinement (AMR) hydrodynamic code, with a linear
Godunov solver and piece-wise linear cell interpolation 
\citep[see][]{Falle:1991}. The entire
computational domain is covered by the two coarsest grids, $G^0$ and
$G^1$. The solution at each position is calculated on all grids that
exist there, and the difference between these solutions is used to
control refinement. 8 grid levels were used in total,
with a factor of 2 refinement between each. In all simulations the 
coarse-grid resolution is equal to the radius of the cloud, with
128 cells per cloud radius on the finest grid.  An advected scalar is used
to distinguish between cloud and ambient material.

The interaction is simplified by adopting a number of assumptions.
The magnetic field is assumed to be too weak to be dynamically
important. We also assume that radiative cooling can be ignored.  This
is valid if the cloud is small enough, and preserves the scale-free
nature of the simulations. The efficiency of thermal conduction in
magnetized turbulent plasmas remains highly uncertain \citep[see][and
references therein]{Pittard:2009}, and for simplicity its effects are
also ignored here. Finally, we also ignore self-gravity. It is
unimportant in a cloud struck by a strong adiabatic shock
\citep{Klein:1994}, but should be included in future work on mildly
supersonic shock cloud interactions.

The turbulence in the sub-grid $k$-$\epsilon$ model is assumed to be
fully developed, and has been calibrated by comparing the computed
growth of shear layers with experiments \citep{Dash:1983}. Turbulent
energy ($k$, per unit mass) is generated by the action of the
turbulent viscosity on the mean flow and is converted to heat at the
dissipation rate per unit mass, $\epsilon$.  Since the turbulent
energy resides mainly in large eddies, while the dissipation occurs in
the small ones, one can think of $k$ and $\epsilon$ as describing the
large-scale and small-scale turbulence respectively. Further details
of the model implementation including the full set of equations that
are solved can be found in \citet{Pittard:2009}.

The ratio of the turbulent and thermal energy densities, $e_{\rm
tb}/e_{\rm th},$ is $\sim 10^{-6}$ in the initial post-shock
flow.  Within the pre-shock medium (including the cloud), we set
$e_{\rm tb}/e_{\rm th} = 0.04$. These values are low enough to not
affect the initial dynamics of the interaction, and are identical to
those used by \citet{Pittard:2009} in their investigation of a Mach 10
shock-cloud interaction (where much higher levels of post-shock
turbulence were also investigated). 

The numerical domain is set large enough so that the cloud is well
dispersed and mixed into the post-shock flow before the shock reaches
its edge.  In this way various global quantities detailed below can be
accurately computed. The shock propagates parallel to the axis of
symmetry, which is the $z$-axis. For clouds with a density contrast
$\chi = 10^{3}$ with respect to the ambient medium, the grid had an
extent of $0 \leq r \leq 24$, with $-910 \leq z \leq 290$ when
$M=1.5$, $-510 \leq z \leq 20$ when $M = 3$, and $-510 \leq z \leq 6$
when $M = 10$, where the unit of length corresponds to a cloud radius,
$r_{\rm c}$. Smaller grids were used when $\chi$ was lower.  All
calculations were performed for an ideal gas with $\gamma=5/3$, and
were scaled so that the fluid variables have values reasonably close
to unity.

Clouds in the ISM do not have infinitely sharp edges, so we adopt the
density profile specified in \citet{Pittard:2009}:
\begin{equation}
\rho(r) = \rho_{\rm amb}[\psi + (1 - \psi)\eta],
\end{equation}
\noindent where
\begin{equation}
\eta = \frac{1}{2}\left(1 + \frac{\alpha - 1}{\alpha + 1}\right),
\end{equation}
$\alpha = {\rm exp}\;\{{\rm min}[20.0,p_{1}((r/r_{\rm c})^{2}-1)]\}$,
and $r$ is the distance from the centre of the cloud. $\psi$ is 
adjusted to obtain a specific density contrast for the
centre of the cloud with respect to the ambient medium ($\chi =
\rho_{\rm max}/\rho_{\rm amb}$). The parameter $p_{1}$ controls the
steepness of the profile at the edge of the cloud.
We set $p_{1}=10$ (i.e. a sharp-edged cloud), 
and place the cloud in pressure equilibrium with its surroundings.  

The evolution of the interaction is studied through various integrated
quantities \citep[see][]{Klein:1994,Nakamura:2006,Pittard:2009}. These
include the effective radii of the cloud normal to and along the axis
of symmetry ($a$ and $c$, respectively), its mass ($m$), its mean
velocity ($\langle v_{\rm z}\rangle$, measured in the frame of the
unshocked cloud), the velocity dispersions in the radial and axial
directions ($\delta v_{\rm r}$ and $\delta v_{\rm z}$, respectively),
its volume ($V$), its mean density ($\langle\rho\rangle$), and the
total circulation produced ($\Sigma$).  The whole of the cloud and the
densest part of its core are distinguished by the value of the scalar
variable $\kappa$ associated with the cloud
\citep[see][]{Pittard:2009}. In this way, each global statistic can be
computed for the region associated only with the core (identified with
the subscript ``core'', e.g., $a_{\rm core}$) or with the entire cloud
(identified with the subscript ``cloud'', e.g., $a_{\rm cloud}$).

The characteristic time for the cloud to be crushed by the shocks driven
into it is the ``cloud crushing'' time, $t_{\rm cc} = \chi^{1/2}
r_{\rm c}/v_{\rm b}$, where $v_{\rm b}$ is the velocity of the shock
in the intercloud (ambient) medium \citep{Klein:1994}. 
Several other timescales are obtained from the simulations. The time for
the average velocity of the cloud relative to that of the postshock
ambient flow to decrease by a factor of 1/$e$ is defined as the ``drag
time'', $t_{\rm drag}$. The ``mixing time'', $t_{\rm mix}$, is defined
as the time when the mass of the core of the cloud, $m_{\rm core}$,
reaches half of its initial value. The zero-point of all time
measurements occurs when the intercloud shock is level with the centre
of the cloud.

The diffusive nature of turbulence means that the cloud gradually
disperses even when no shock is present. The time needed for the
maximum density in the core of a cloud with $\chi=10$ to drop to half
its original value is $\approx 150\,t_{\rm cc}$, reducing to $\approx
4\,t_{\rm cc}$ for a cloud with $\chi=10^{3}$. A comparison of these
timescales to the cloud destruction timescales discussed in
Section~\ref{sec:results} reveals that the evolution of the cloud is
always dominated by its interaction with the shock.

\section{Stages and ``Mach scaling''}
\label{sec:stages}
The main stages of an adiabatic, non-magnetized and non-conducting
interaction of a high Mach number shock with a cloud have been
described many times in the literature \citep[see,
e.g.,][]{Klein:1994,Pittard:2009}. Initially the cloud is compressed
mainly in the $z$-direction by the incident shock which propagates
into the cloud and by a shock driven into the back of the cloud, and a
bow shock propagates upstream into the ambient medium\footnote{The
Mach numbers of the reflected and transmitted shocks, in the
one-dimensional planar limit, are calculated in
\citet{Miesch:1994}. Unfortunately, in the strong, adiabatic shock
limit, the analytical solution for the value of the Mach number of the
reflected shock, $M_{\rm r}$, does not satisfy the original assumption
that $M_{\rm r}^{2} \gg 1$, and so is known to be quantitatively
incorrect.  A \mbox{6$^{\rm th}$-order} polynomial must instead be
solved numerically.}. The over-pressured cloud then expands downstream
and laterally, and Rayleigh-Taylor (RT) and Kelvin-Helmholtz (KH)
instabilities destroy the cloud and mix its material into the
surrounding flow. The cloud is more rigid, and is better able to
resist the passage of the shock, as $\chi$ increases\footnote{In fact,
only clouds with $\chi > 10^{3}$ should be considered ``rigid''
\citep{Miesch:1994}.}.

The conditions behind a strong shock are virtually independent of the
sound speed ahead of the shock (e.g., for $M=10$, the post-shock
density is within a few percent of its value at $M \rightarrow
\infty$, while the normalized post-shock pressure is within a fraction
of a percent).  Since the inviscid, adiabatic, non-magnetized and
non-conducting hydrodynamic equations are invariant under the
transformation
\begin{equation}
\label{eq:trans}
t \rightarrow tM, \hspace{10mm} v \rightarrow \frac{v}{M}, \hspace{10mm} P \rightarrow \frac{P}{M^{2}},
\end{equation}
with the position and density unchanged, the time evolution of the
cloud is independent of the Mach number of the shock when expressed in
units of $t/t_{\rm cc} \propto tM$ in the limit $M \rightarrow
\infty$. This is referred to as ``Mach scaling'', and was demonstrated
to be reasonably valid for clouds with sharp \citep{Klein:1994} and
with smooth \citep{Nakamura:2006} boundaries.

The system of equations used for the $k$-$\epsilon$ calculations are
shown in \citet{Pittard:2009}.  They are invariant under the additional
transformation
\begin{equation}
\label{eq:trans2}
k \rightarrow \frac{k}{M^{2}}, \hspace{10mm} \epsilon \rightarrow \frac{\epsilon}{M^{3}},
\end{equation}
so Mach scaling occurs in this case also when $M \rightarrow
\infty$. 

At lower Mach numbers, the post-shock conditions are dependent on the
shock Mach number and Mach scaling is not applicable. The evolution of
the interaction is also different to the strong-shock case, as shown
below.

\section{Results}
\label{sec:results}
In this section we first examine the Mach number dependence of the
interaction of a shock with a cloud of high density contrast
($\chi=10^{3}$).  This extends the work in \citet{Pittard:2009} to a
range of Mach numbers. The dependence of the interaction on $\chi$ is
then examined, after which we are able to draw some conclusions about
the necessary conditions for material stripped from the cloud to form
identifiable tails. We then examine three main themes: the Mach
scaling of the interaction and the principal differences which result
when the post-shock flow is subsonic; the generation of turbulence in
the interaction; and the mass-loss rate and the lifetime of the cloud.
Some key parameters from the simulations are tabulated in
Table~\ref{tab:resultslo}.

\subsection{Cloud morphology and turbulence}
\subsubsection{Interactions with $\chi=10^{3}$}
Fig.~\ref{fig:m1.5lokeps} shows snapshots of the density
distribution at different times for the case of an adiabatic Mach 1.5
shock impacting a cloud with a density contrast $\chi=10^{3}$. The
interaction is much milder than in the Mach 10 case
\citep[see][]{Pittard:2009}, which results in several major
differences \citep[cf.][]{Nakamura:2006}: i) the postshock flow is
subsonic with respect to the cloud, so a bowwave rather than a
bowshock forms ahead of the cloud; ii) the compression of the cloud is
more isotropic; iii) a strong vortex ring is not produced; iv) the
smaller velocity difference at the slip surface around the cloud
limits the KH and RT instabilities which develop strongly in the
$M=10$ case; v) it takes much longer for the cloud to be mixed
into the surrounding flow and for it to accelerate to the intercloud
postshock speed.  A comparison with the results from
an inviscid code (without the subgrid turbulence model) reveals that
the evolution is overall very similar.  This was also the case for the
Mach 10 interaction studied in \citet[][]{Pittard:2009},
though significant differences
occur if the post-shock medium sweeping over the cloud is very
turbulent.

Direct comparison of the density structure at $t=5.66\;t_{\rm cc}$
for interactions with Mach 1.5, 2, 3, 6 and 10 shocks are made in
Fig.~\ref{fig:machcomplo}.  There are significant differences between
the density distributions from the low Mach number interactions on the
one hand, and the high Mach number interactions on the other. At high
Mach numbers ($M > 2.76$) the post-shock flow is supersonic, and a
bowshock develops around the cloud. The shock driven into the cloud is
stronger, and less symmetric.  The faster flow speed past the cloud
results in faster growth of RT and KH instabilities, and more rapid
acceleration of the cloud.

The $M=6$ and $M=10$ simulations show dense elongated tails at
$t=5.66\,t_{\rm cc}$, whereas the lower Mach number simulations do
not. Since long tails do eventually form even in the Mach 1.5
interaction, it is apparent that material stripped off the cloud forms
a longer tail at a given time (measured in units of $t_{\rm cc}$) as
the Mach number increases (the $M=3$ simulation forms a long
well-defined tail by $t=8.54\;t_{\rm cc}$). This is because the
stripping of material is more efficient at higher Mach numbers due to
the faster growth of RT and KH instabilities.  Fig.~\ref{fig:machcomplo}
also shows that the acceleration of the cloud increases with the Mach
number of the interaction, though at very high Mach numbers Mach
scaling (Section~\ref{sec:stages}) holds.

The development of turbulence in model m1.5c3 is shown in
Fig.~\ref{fig:m1.5tk}, where the turbulent energy per unit mass, $k$,
is displayed. $k$ is initially created in a thin turbulent boundary
layer at the surface of the cloud where a region of high shear
exists. The turbulent eddies are then advected by the flow and a
turbulent wake develops downstream of the cloud, with a setup time
$\sim t_{\rm cc}$.  This behaviour is qualitatively similar to the
Mach 10 case shown in \citet{Pittard:2009}, although the peak
turbulent intensity is smaller, and a supersonic vortex ring and its
associated turbulence are not present.

\subsubsection{Interactions with $\chi=10^{2}$ and $\chi=10$}
\label{sec:chi10and100}
Clouds with density contrast $\chi=10^{2}$ are less rigid obstacles
to the passage of a shock than clouds with $\chi=10^{3}$. This can
easily be discerned by comparing snapshots of the mass density at
$t=4.85\;t_{\rm cc}$ in interactions of shocks of various Mach numbers
with clouds of density contrast $\chi=10^{2}$ as shown in
Fig.~\ref{fig:dens1e2machcomplo}, to the snapshots in
Fig.~\ref{fig:machcomplo} of interactions with $\chi=10^{3}$.

\begin{figure*}
%8.5cm for 2-col layout, 5.7cm for 3-col layout
\psfig{figure=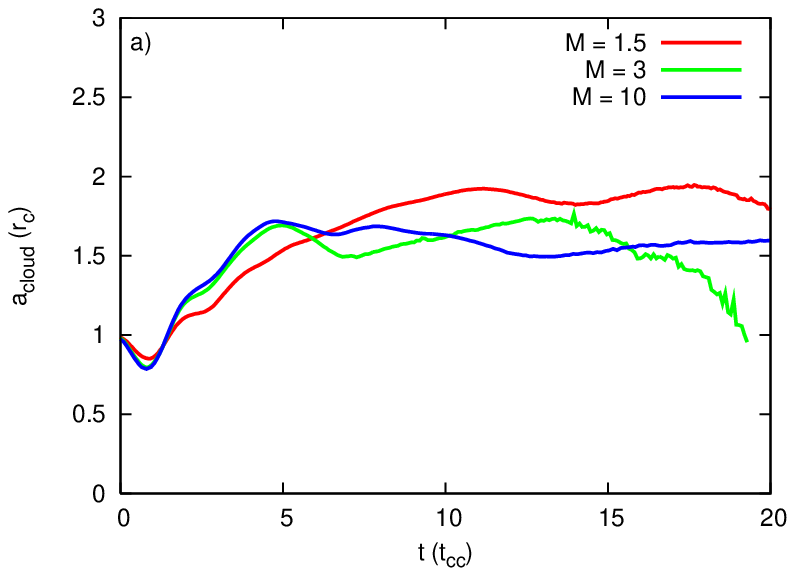,width=5.7cm}
\psfig{figure=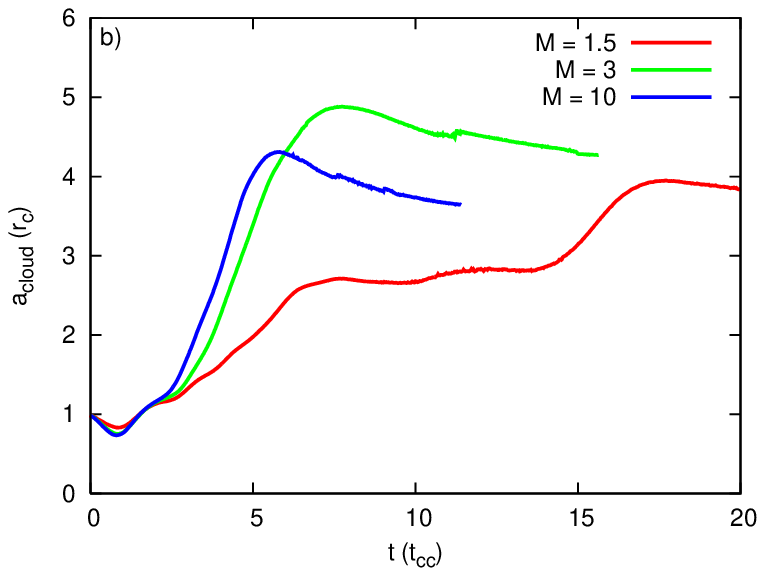,width=5.7cm}
\psfig{figure=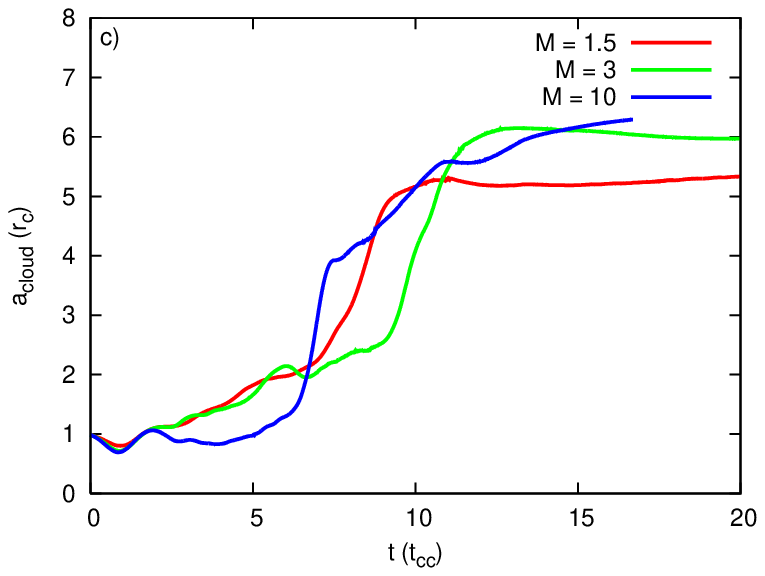,width=5.7cm}
\caption[]{Time evolution of the effective transverse radius of the
cloud, $a_{\rm cloud}$, for various Mach numbers and density contrasts:
(a) $\chi=10$, (b) $\chi=10^{2}$, (c) $\chi=10^{3}$.}
\label{fig:machcomp_cloud_shape}
\end{figure*}

In fact, \citet{Klein:1994} noted that the {\em nature} of the
interaction appears to change with $\chi$: at low $\chi$ the cloud is
liable to break up into several large fragments, whereas at large
$\chi$ the core remains more intact and material is stripped from its
surface (this is a less effective mixing mechanism). This observation
is partially supported by the density snapshots shown in
\citet{Pittard:2009}, where the last panel in each of Figs.~9, 10 and
4 shows snapshots at similar times for a Mach 10 shock interacting
with a cloud where $\chi=10$, $10^{2}$, and $10^{3}$,
respectively. However, a closer examination of the statistics of the
interaction reveals that in almost all of the cases investigated in
this work, the $\chi=10^{2}$ cloud is destroyed faster than clouds
with $\chi=10$ and $10^{3}$ (the exception is the $M=1.5$ simulation,
where the $\chi=10$ cloud is actually destroyed faster). The generally
rapid destruction of clouds with $\chi=10^{2}$ appears to be due to
the fact that such clouds undergo a more rapid increase in their
transverse radius as they expand following the initial compression
caused by the passage of the shock (see
Fig.~\ref{fig:machcomp_cloud_shape}). This makes the cloud more
vulnerable to fragmentation.

The interaction of an adiabatic shock with a cloud of density contrast
$\chi=10$ has been discussed many times in the literature, which in
the interest of brevity we do not repeat here. However,
\cite{Klein:1994} notes that the axial stretching of the cloud
increases with $\chi$. This is important insofar as we do not see a
well-defined tail in any of our simulations when $\chi \ltsimm
10^{2}$.  It appears, therefore, that well-defined tails only form in
interactions with density contrasts $\chi \gtsimm 10^{3}$, though in
such cases they can form across a wide range of Mach numbers.

\begin{figure*}
%8.5cm for 2-col layout, 5.7cm for 3-col layout
\psfig{figure=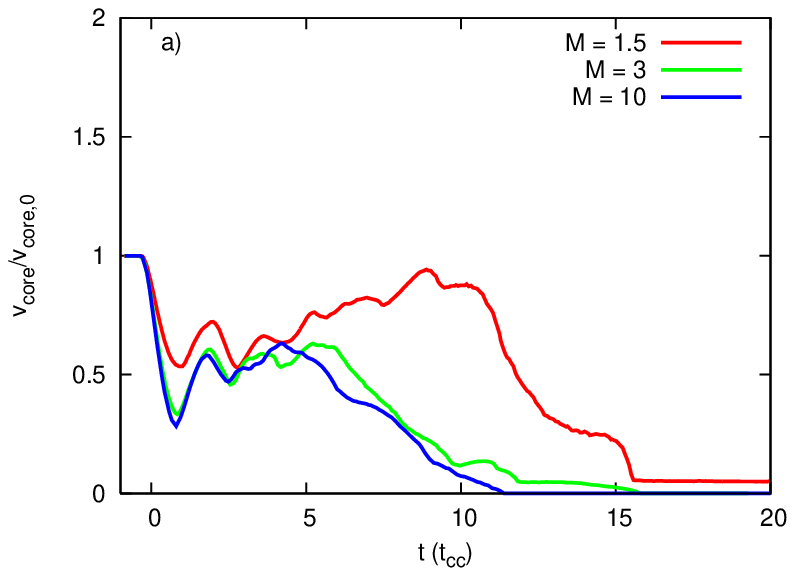,width=5.7cm}
\psfig{figure=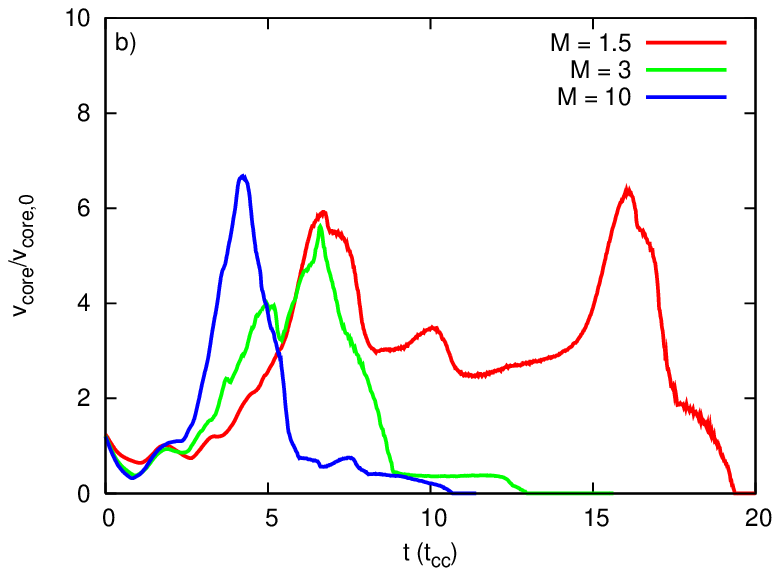,width=5.7cm}
\psfig{figure=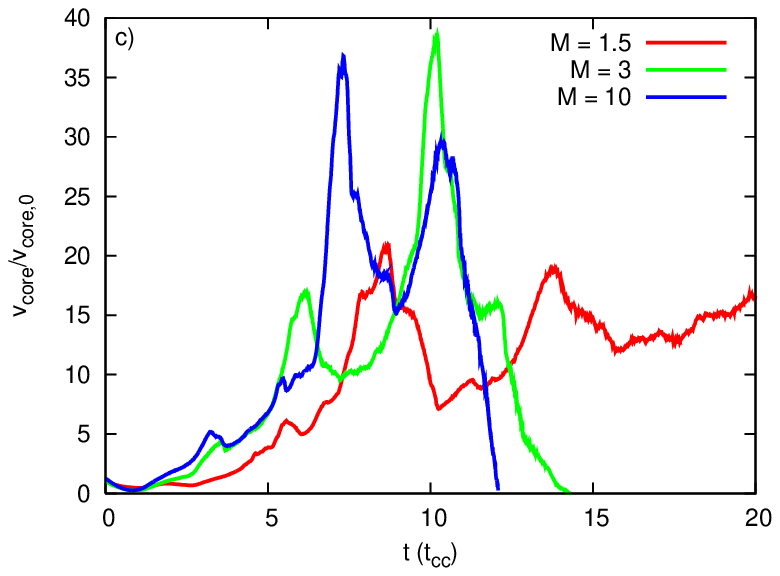,width=5.7cm}
\caption[]{Time evolution of the volume of the core of the
cloud, $V_{\rm core}$, normalized to its initial value, 
for various Mach numbers and density contrasts:
(a) $\chi=10$, (b) $\chi=10^{2}$, (c) $\chi=10^{3}$.}
\label{fig:machcomp_volume}
\end{figure*}

\begin{figure*}
%8.5cm for 2-col layout, 5.7cm for 3-col layout
\psfig{figure=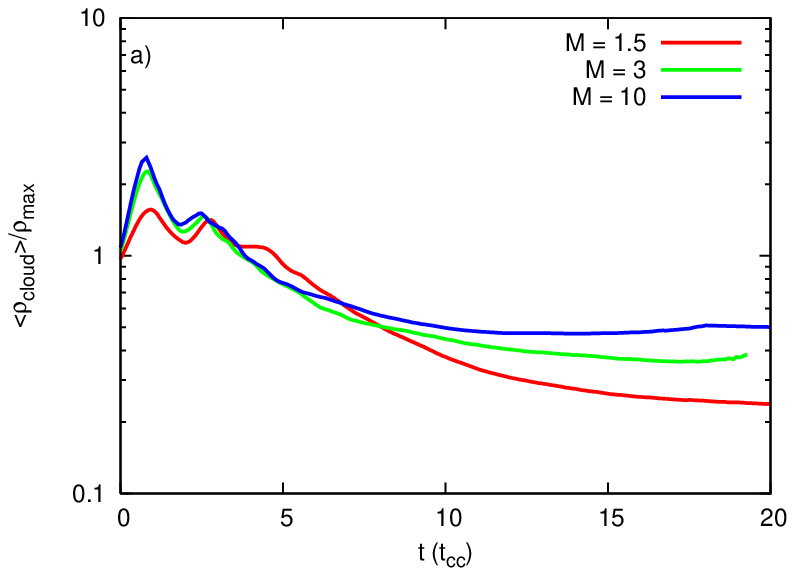,width=5.7cm}
\psfig{figure=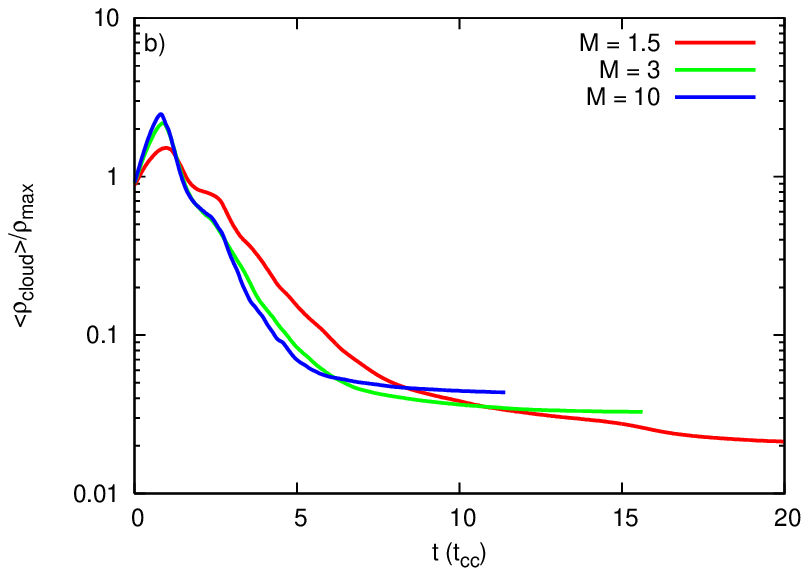,width=5.7cm}
\psfig{figure=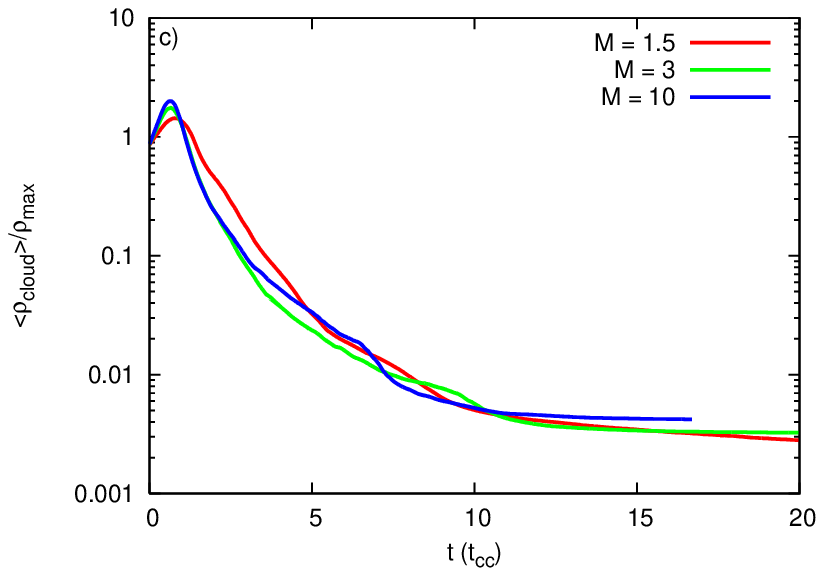,width=5.7cm}
\psfig{figure=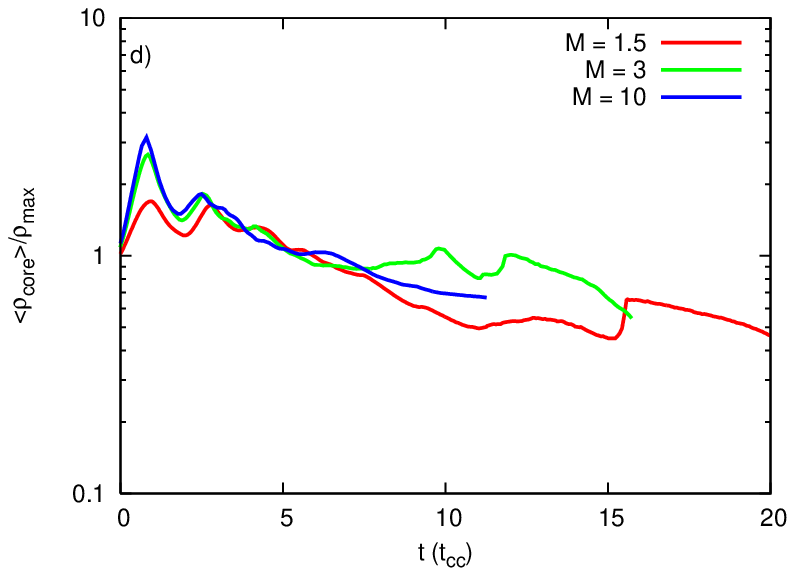,width=5.7cm}
\psfig{figure=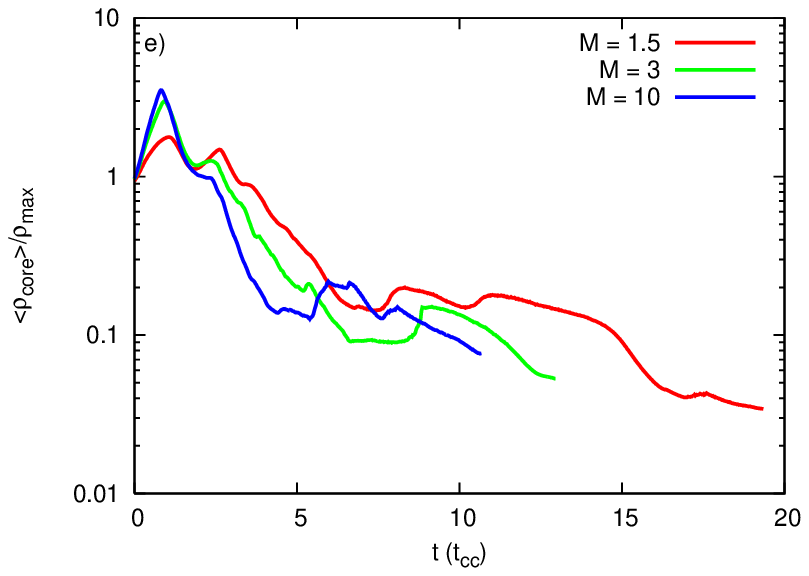,width=5.7cm}
\psfig{figure=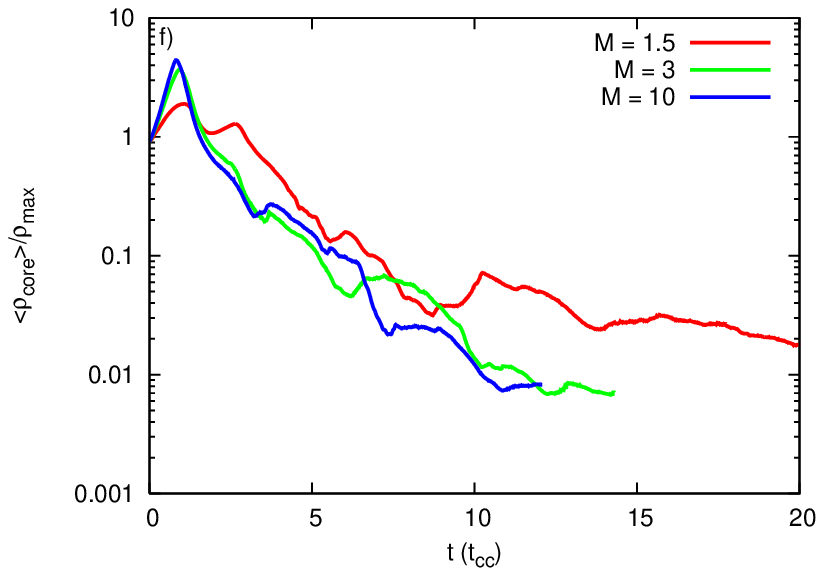,width=5.7cm}
\caption[]{Top panels: Time evolution of the mean density of the cloud,
$\langle\rho_{\rm cloud}\rangle$, normalized to the initial maximum cloud density,
for various Mach numbers and density contrasts:
(a) $\chi=10$, (b) $\chi=10^{2}$, (c) $\chi=10^{3}$. Bottom panels: As top,
but for the mean density of the core, $\langle\rho_{\rm core}\rangle$.}
\label{fig:machcomp_density}
\end{figure*}

\begin{figure*}
\psfig{figure=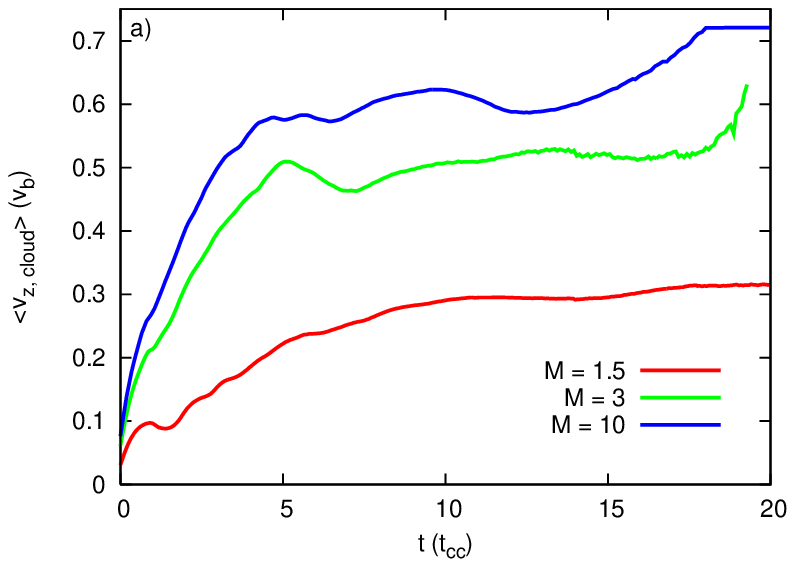,width=5.7cm}
\psfig{figure=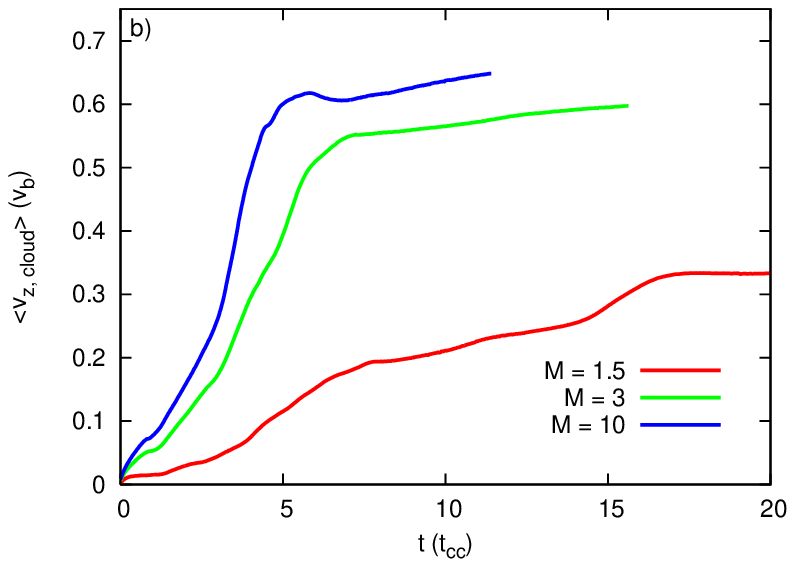,width=5.7cm}
\psfig{figure=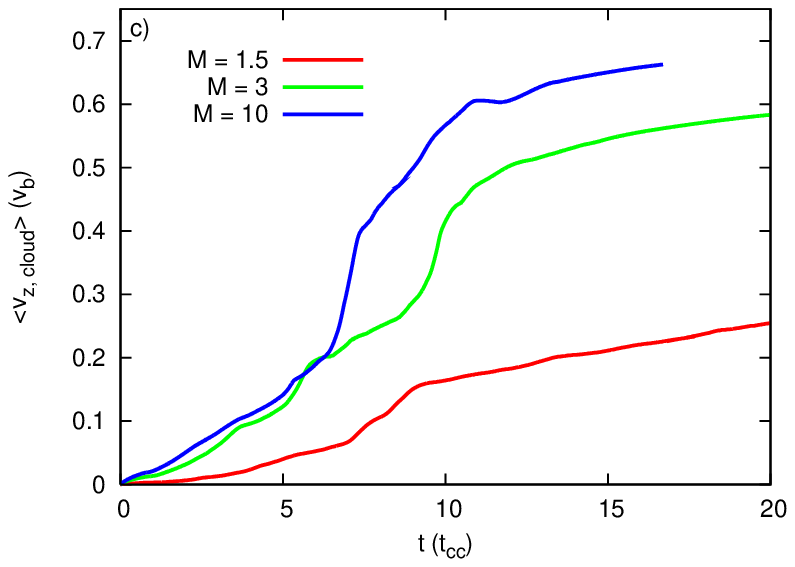,width=5.7cm}
\psfig{figure=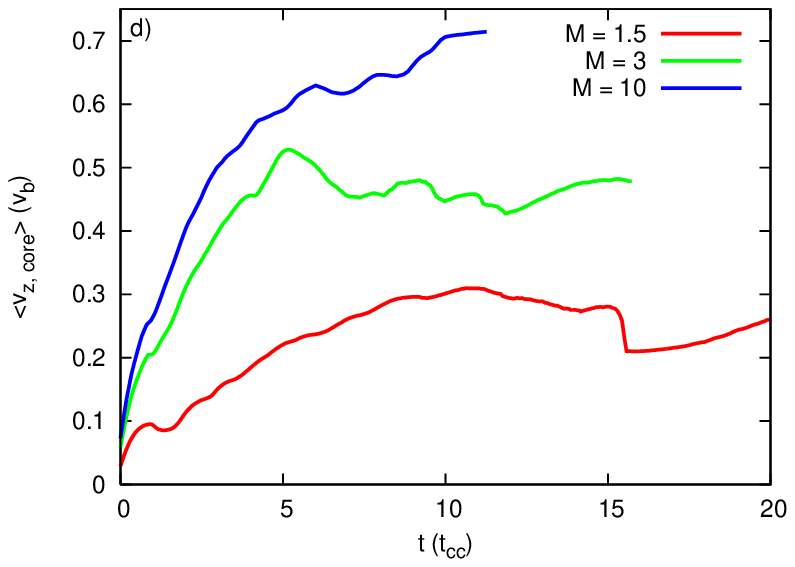,width=5.7cm}
\psfig{figure=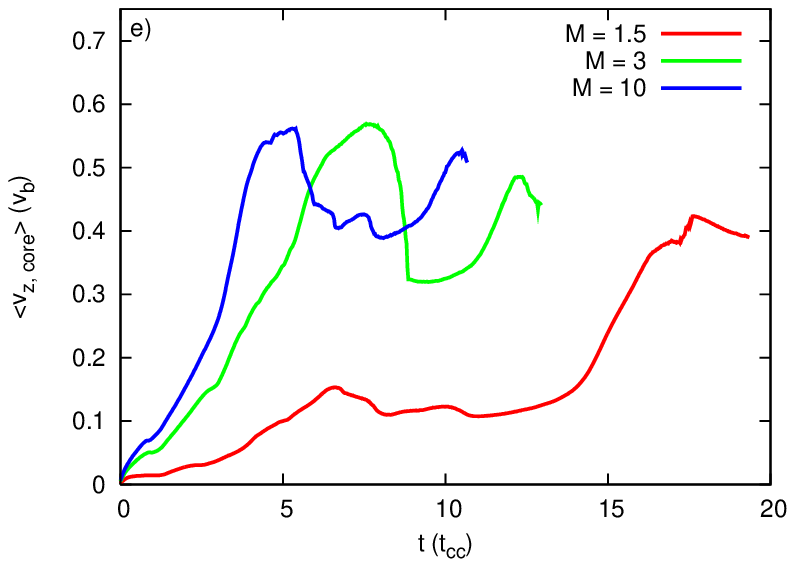,width=5.7cm}
\psfig{figure=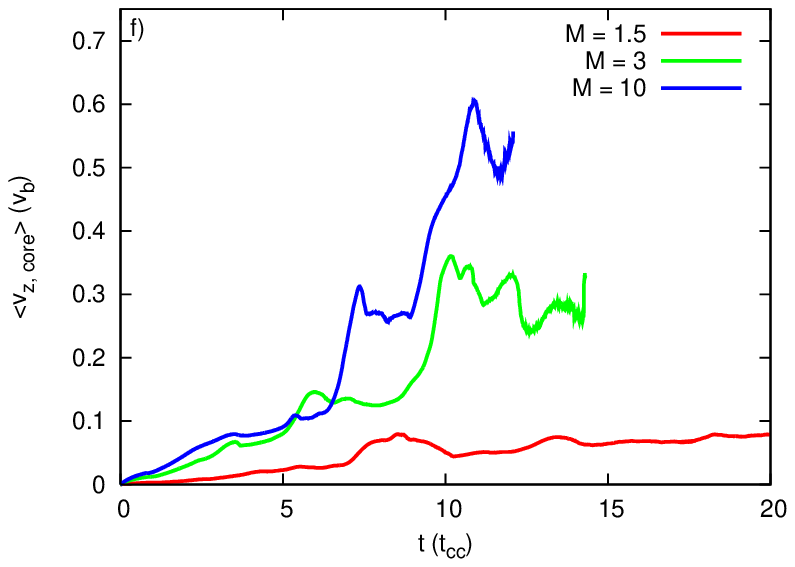,width=5.7cm}
\caption[]{Top panels: Time evolution of the cloud mean velocity,
$\langle v_{\rm z} \rangle_{\rm cloud}$, for various Mach numbers and
density contrasts: (a) $\chi=10$, (b) $\chi=10^{2}$, (c)
$\chi=10^{3}$. Bottom panels: as the top panels, but for the core mean
velocity, $\langle v_{\rm z} \rangle_{\rm core}$.}
\label{fig:machcomp_cloud_meanvelocity}
\end{figure*}

\begin{figure*}
\psfig{figure=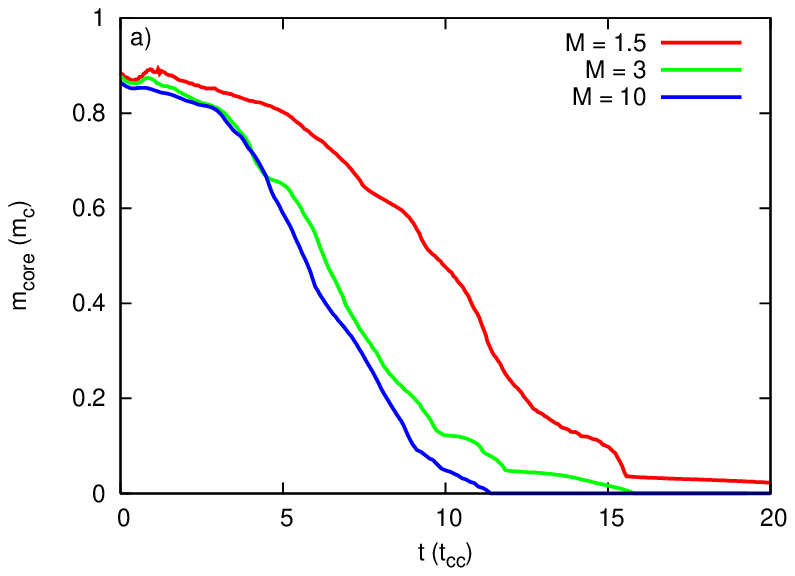,width=5.7cm}
\psfig{figure=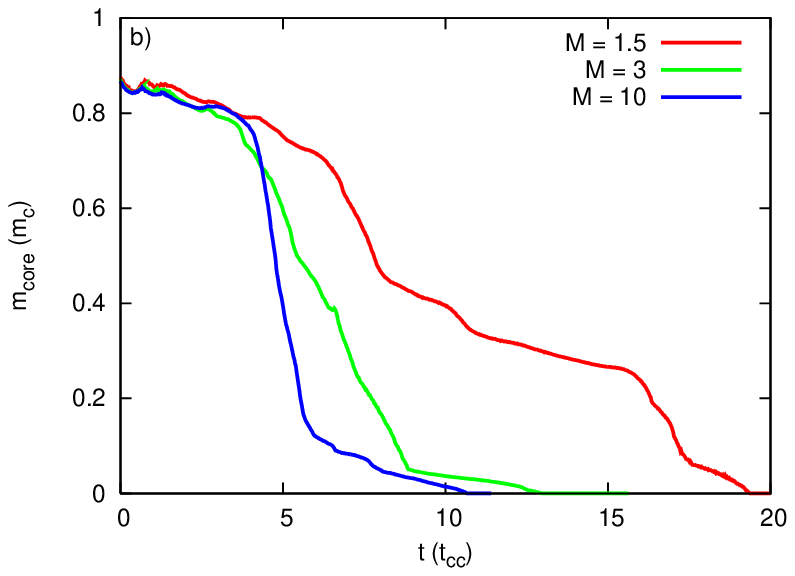,width=5.7cm}
\psfig{figure=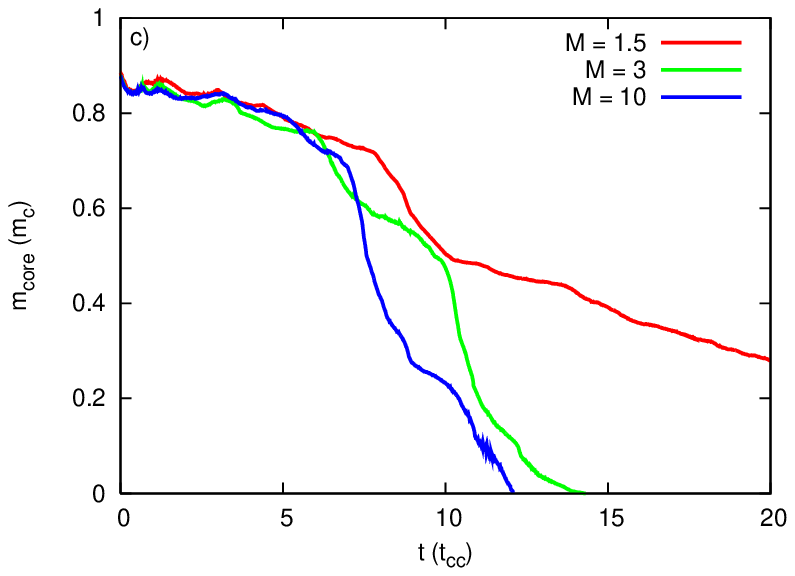,width=5.7cm}
\caption[]{Time evolution of the core mass, $m_{\rm core}$, for various Mach
numbers and density contrasts: (a) $\chi=10$, (b) $\chi=10^{2}$,
(c) $\chi=10^{3}$.} 
\label{fig:machcomp_core_mass}
\end{figure*}

\begin{figure*}
\psfig{figure=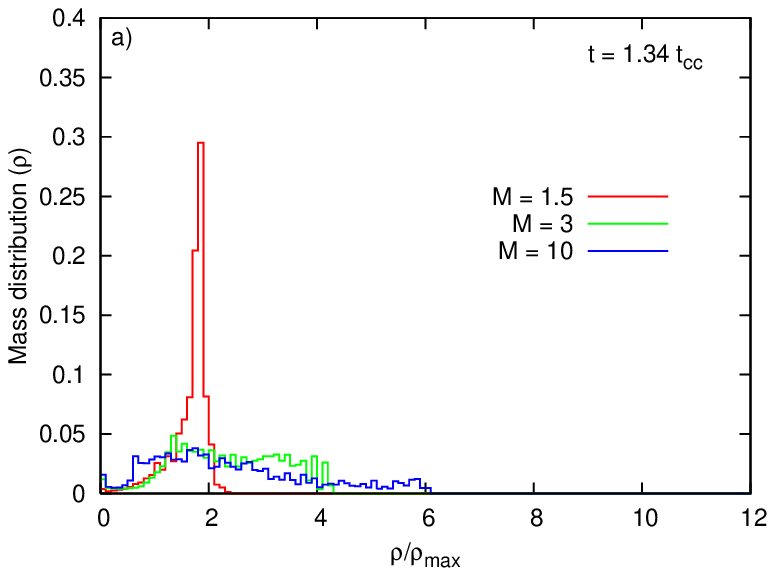,width=5.7cm}
\psfig{figure=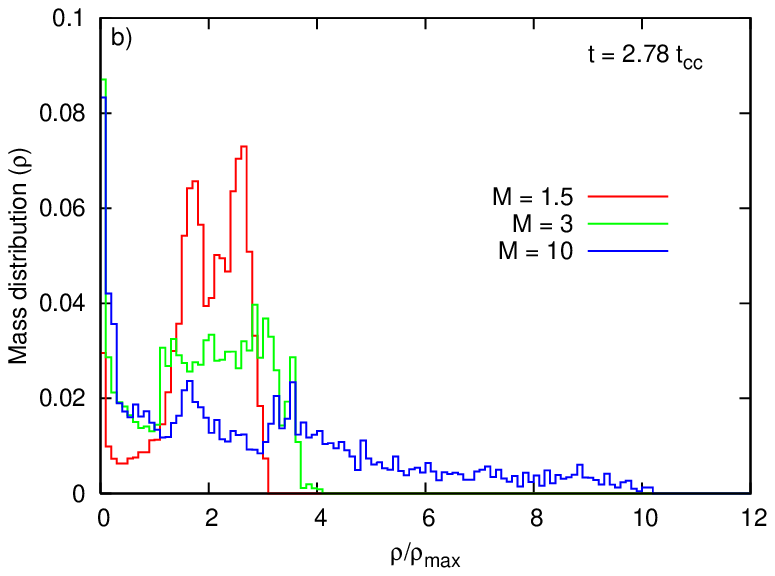,width=5.7cm}
\psfig{figure=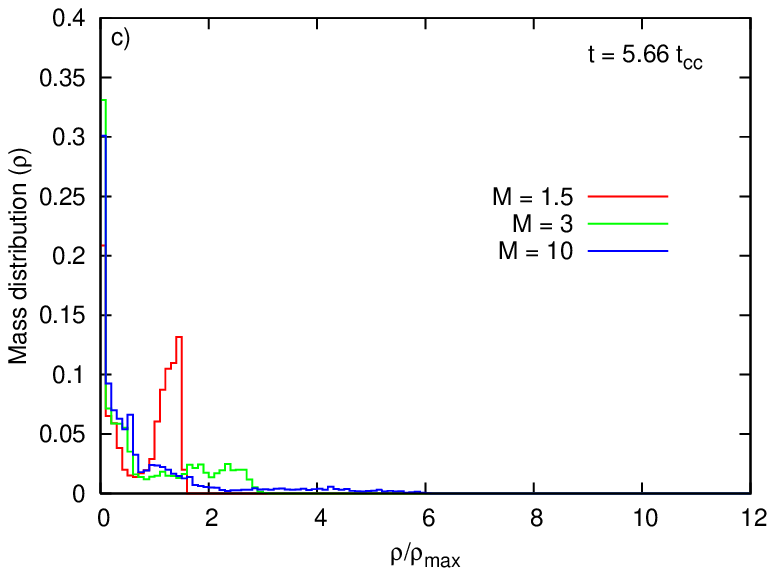,width=5.7cm}
\caption[]{Time evolution of the mass density distributions for models with $\chi=10^{3}$ and
varying shock Mach numbers. The histograms indicate the
cloud mass contained within each density bin of width $0.1
\rho_{\rm max}$, normalized to the total cloud mass, $m_{\rm c}$. From
left to right the mass fractions were computed at $t=1.34, 2.78$ and
$5.66\;t_{\rm cc}$.}
\label{fig:massfrac}
\end{figure*}

\begin{figure*}
\psfig{figure=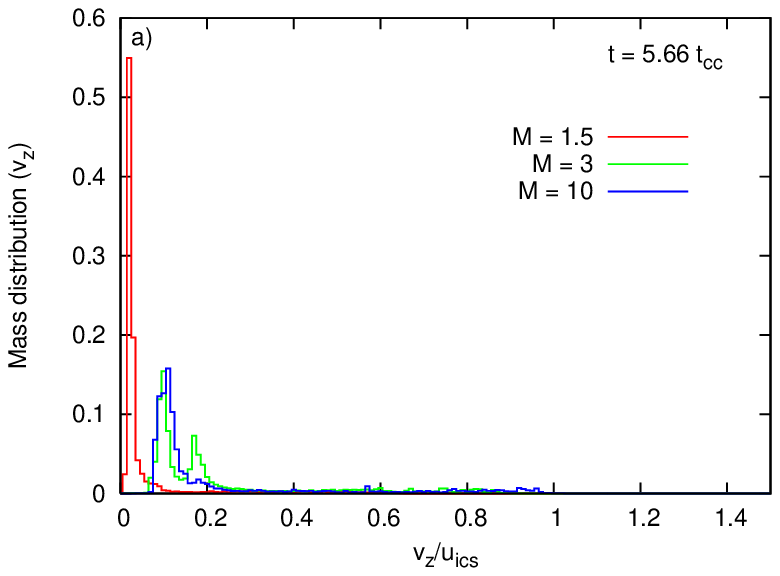,width=5.7cm}
\psfig{figure=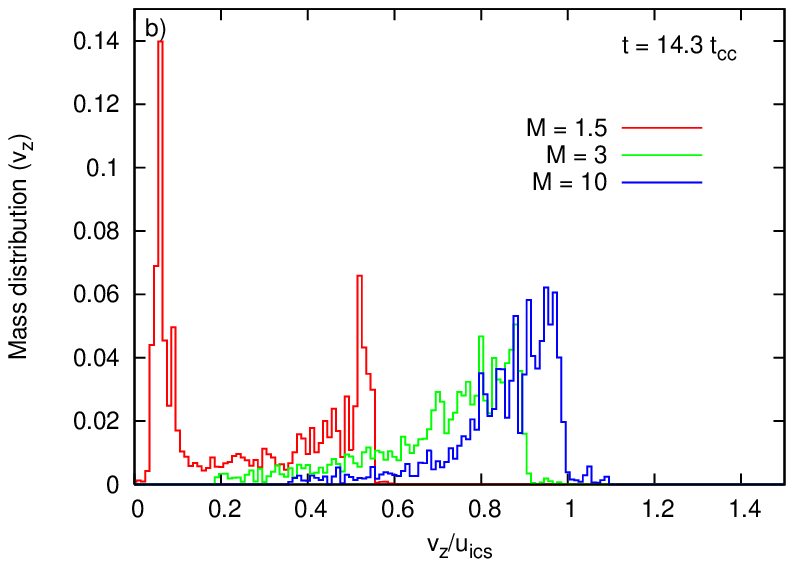,width=5.7cm}
\psfig{figure=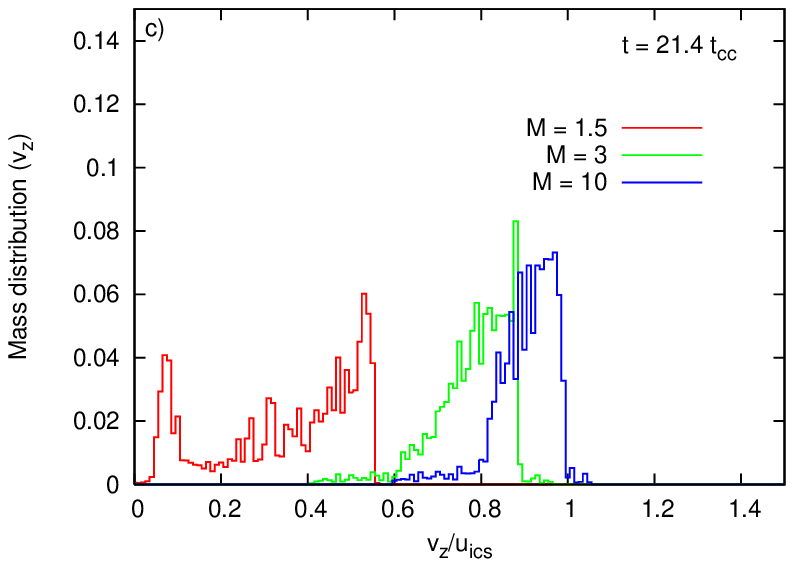,width=5.7cm}
\caption[]{Mass distributions as a function of $v_{\rm z}$ (normalized to the post-shock
velocity in the ambient medium, $u_{\rm ics}$) for models with
$\chi=10^{3}$ and varying shock Mach numbers at $t = 5.66, 14.3$ and
$21.4\;t_{\rm cc}$.  The histograms denote the mass contained within a
velocity bin of width $0.01\;u_{\rm ics}$ for a line of sight parallel
to the z-axis. The integrated mass is the mass of the cloud, $m_{\rm
c}$.}
\label{fig:mass_spec}
\end{figure*}

\begin{figure*}
\psfig{figure=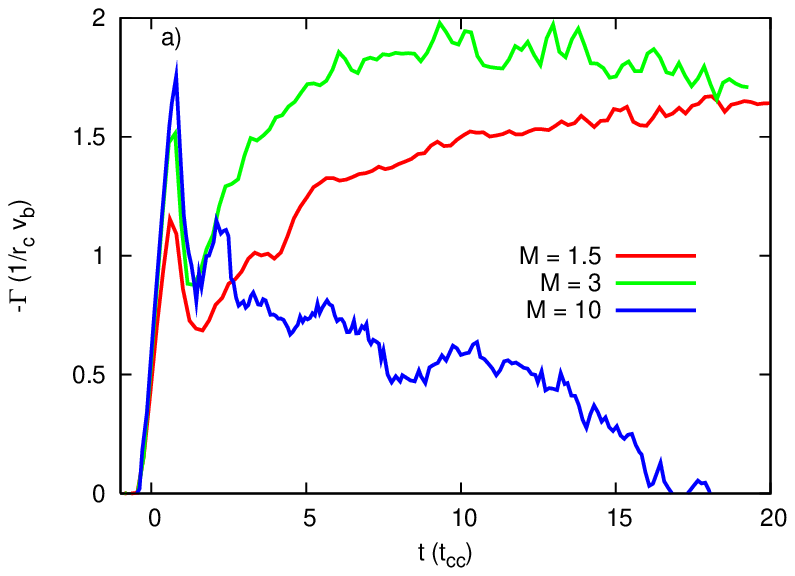,width=5.7cm}
\psfig{figure=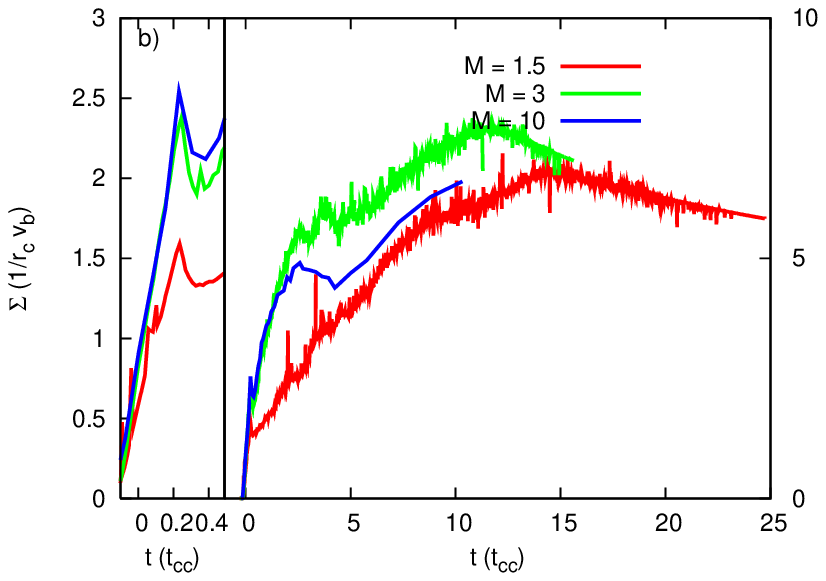,width=5.7cm}
\psfig{figure=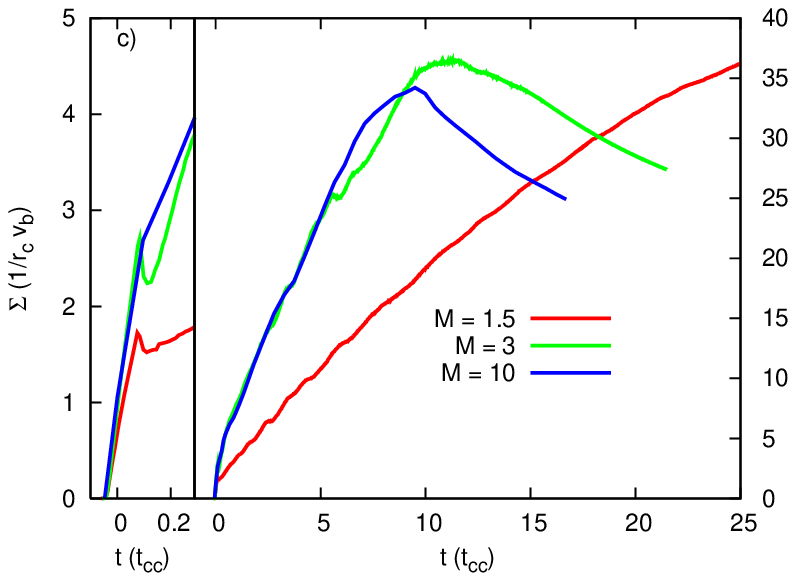,width=5.7cm}
\caption[]{Time evolution of the circulation, $\Gamma_{\rm circ}$, for
various Mach numbers and density contrasts: (a) $\chi=10$, (b)
$\chi=10^{2}$, (c) $\chi=10^{3}$.}
\label{fig:machcomp_circulation}
\end{figure*}

\subsection{Mach scaling and principle changes for subsonic post-shock flow}
In this section we further examine how the interaction changes with
the Mach number, and the major differences which occur when the
post-shock flow is subsonic.

\subsubsection{Cloud shape, volume, and mean density}
\label{sec:shape}
Fig.~\ref{fig:machcomp_cloud_shape} shows that the initial
compression of the cloud in the transverse direction is weakest in the
$M=1.5$ simulation, but comparable in the $M=3$ and $M=10$
simulations. The transverse expansion which follows as the cloud seeks
to re-establish pressure equilibrium is typically slower in the
$M=1.5$ simulations, but occurs at a similar rate in the $M=3$ and
$M=10$ simulations.

That the initial compression of the cloud is weaker at low Mach
numbers can also be discerned from Fig.~\ref{fig:machcomp_volume}(a).
For low density contrasts ($\chi \ltsimm 10$) the core of the cloud
maintains a larger volume than when hit by a stronger shock. However,
clouds with higher density contrast hit by a strong shock undergo a
violent re-expansion after their initial compression, causing their
core volume to exceed their counterparts hit by weaker shocks (see
panels (b) and (c) in Fig.~\ref{fig:machcomp_volume}). At later times
the volume of the cloud core decreases as the core material is
gradually ablated and mixed into the surrounding flow.

The reduced compressions behind low Mach number shocks relative to
their stronger counterparts cause lower peak densities of the clouds
and their core material, $\langle \rho_{\rm cloud}\rangle/\rho_{\rm
max}$ and $\langle \rho_{\rm core}\rangle/\rho_{\rm max}$,
respectively (see Fig.~\ref{fig:machcomp_density}).  And although the
decline in these quantities over the period $t \ltsimm 5\;t_{\rm cc}$
is slower in the Mach 1.5 simulations, due to the slower stripping of
material into the flow, the final value of $\langle \rho_{\rm
cloud}\rangle/\rho_{\rm max}$ that is reached is lower in the Mach 1.5
simulations than in the higher Mach number simulations, since the
density of the postshock ambient material is lower in this case. This
is particularly noticeable in Fig.~\ref{fig:machcomp_density}(a).

\subsubsection{Cloud velocity}
At lower Mach numbers there is a gentler acceleration of the cloud and
its core in the axial direction, shown by the mean velocity
statistics, $\langle v_{\rm z,cloud}\rangle$ and $\langle v_{\rm
z,core}\rangle$, respectively (panels (a-c) and (d-e) in
Fig.~\ref{fig:machcomp_cloud_meanvelocity}). This reflects both the
weaker shock that is initially driven into the cloud, and the slower
speed of the post-shock flow past the cloud. The acceleration of the
cloud to a velocity of $1/e$ times the ambient postshock flow speed is
discussed further in Section~\ref{sec:global_tdrag_tmix}.

\subsubsection{Cloud mass}
\label{sec:cloud_mass}
The slower post-shock flow in the simulations with lower Mach numbers
affects the growth rate of RT and KH instabilities.  This in turn
affects the speed at which material is ripped from the surface of the
cloud and the time for the cloud to mix into the flow. Hence, $m_{\rm
core}$ declines more rapidly in simulations with higher Mach numbers
(see Fig.~\ref{fig:machcomp_core_mass}). In high Mach number
interactions the core completely disappears shortly after $t_{\rm
mix}$, whereas at lower Mach numbers material may be identified as
originating from the core for times considerably after $t_{\rm mix}$.
Also of note is that the slope of $m_{\rm core}$ in the $M=1.5$
simulations appears to get shallower with increasing $\chi$
(Fig.~\ref{fig:machcomp_core_mass}). The maximum slopes of the $M=3$
and $M=10$ models are particularly steep when $\chi=10^{2}$, which is
likely related to the rapid transverse expansion of the cloud radius,
and subsequent strong mixing, taking place in these models (see
Section~\ref{sec:chi10and100}).

The evolution of the cloud and the way that its material mixes with
the surrounding gas can also be studied via its mass distribution in
density or velocity space. Figs.~\ref{fig:massfrac} and~\ref{fig:mass_spec}
show this evolution for clouds with $\chi=10^{3}$.
Fig.~\ref{fig:massfrac} shows the fraction
of cloud mass over a range of density bins of width $0.1\;\rho_{\rm
max}$. The first subfigure is shown for $t=1.34\;t_{\rm cc}$, when the 
transmitted shock has swept through
the entire cloud (see, e.g., Fig.~\ref{fig:m1.5lokeps}).  In the
$M=1.5$ simulation, the weak shock transmitted into the cloud
initially compresses the gas by a factor of 1.7. Slightly higher
values are subsequently obtained as the shock converges on the cloud
centre.  In contrast, the stronger shocks in the $M=3$ and $M=10$
simulations produce higher compressions which can exceed a factor of
10 \citep[see Fig.~19 in][]{Pittard:2009}. At later times the
density of the cloud material drops slowly towards the ambient
post-shock density as the cloud continues to expand into and mix with
the surrounding gas.

Fig.~\ref{fig:massfrac} also reveals that flatter mass distributions
are obtained during higher Mach number interactions, whereas the
distribution is more concentrated and peaky in simulations with lower
Mach numbers.  Two distinct clusters of mass at densities above and
below $2\;\rho_{\rm max}$ can be seen in the $M=1.5$ simulation at
$t=2.78\;t_{\rm cc}$.  This is not just a consequence of the higher
Mach number cases being more evolved - the mass distribution in the
$M=1.5$ simulation never approaches the flatness seen in the higher
Mach number cases, even at much later times.

The flatter mass distribution in the high Mach number cases reflects
differences in the dynamical state of the cloud core with the Mach
number of the interaction.  At $t\approx 1.34\,t_{\rm cc}$, the cloud
hit by the Mach 10 shock is violently re-expanding \citep[see Fig.\,4
of][]{Pittard:2009}. This expansion is supersonic, which leads to a
relatively low density in the central region of the cloud, with higher
densities in the shocked region around the periphery of the cloud.  At
$t\approx 2.78\,t_{\rm cc}$, the cloud core is relatively dispersed,
with fingers of somewhat lower density material around its
edges. Therefore, at both of these instances, the density of material
in the core of the cloud spans a wide range.  In contrast, in the Mach
1.5 case, Fig.~\ref{fig:m1.5lokeps} shows that at $t=1.34\,t_{\rm
cc}$, the core density of the cloud is approximately constant.  Since
the interaction is much milder, the gentler rebounding of the cloud in
the Mach 1.5 case does not lead to such a wide range of densities, and
therefore a more concentrated and peaky distribution is obtained as
seen in Fig.~\ref{fig:massfrac}.

Fig.~\ref{fig:mass_spec} shows the mass distribution function in
velocity space within the entire cloud integrated along the $z$-axis.
The histograms indicate the fraction of the cloud mass contained
within a velocity bin of width $0.01\,u_{\rm ics}$, where $u_{\rm
ics}$ is the post-shock velocity of the ambient medium. Initially the
cloud is at rest, and its velocity is zero. When the material within
the cloud is fully mixed into the surrounding flow its velocity is
equal to $u_{\rm ics}$ (a small fraction of mass exceeds this value
due to turbulent motions). The slower acceleration of material in the
$M=1.5$ simulation is again clearly evident.

\subsubsection{Circulation}
\label{sec:global_circ}
A central aspect of the interaction of a shock with a cloud is the
development of vorticity, ${\mathbf \omega}$, and circulation, $\Gamma$
\citep[see][]{Klein:1994}. Production of the 
latter can be divided into three main
components: at the interface
between the cloud and the surrounding flow by the initial
passage of the shock ($\Gamma_{\rm shock}$) and the subsequent
postshock flow ($\Gamma_{\rm post}$), and at 
the triple points associated with the Mach-reflected
shocks behind the cloud ($\Gamma_{\rm ring}$).
For the general case of
a shock of Mach number $M$ and a ratio of specific heats $\gamma$,
these components are:
\begin{eqnarray}
\label{eq:gamma_shock}
\Gamma_{\rm shock} & \approx & - \frac{6}{\gamma + 1}\left(\frac{M^2 - 1}{M^2}\right) \left(1 - \frac{1}{\chi^{1/2}}\right) v_{\rm b}r_{\rm c},\\ 
\label{eq:gamma_post}
\Gamma_{\rm post} & \approx & - \frac{1}{(\gamma + 1)^2}\left(\frac{M^2 - 1}{M^2}\right)^2  \left(\frac{\chi^{1/2} t_{\rm drag}}{t_{\rm cc}}\right)v_{\rm b}r_{\rm c},\\ 
\label{eq:gamma_ring}
\Gamma_{\rm ring} & \approx & \frac{2}{\gamma+1}\left(\frac{M^2 - 1}{M^2}\right) v_{\rm b}r_{\rm c}. 
\end{eqnarray}
Note that the circulation generated by the vortex ring is positive,
while the passage of the shock and the postshock flow generates
negative circulation.

The contribution of these three components to the total circulation
can be seen in Fig.~\ref{fig:machcomp_circulation}(a). The passage of
the shock over the cloud causes the initial rise to maximum
($\Gamma_{\rm shock}$), with the formation of the vortex ring behind
the cloud ($\Gamma_{\rm ring}$) causing the subsequent drop. The
post-shock flow past the cloud generates further vorticity which
increases the circulation after this minimum ($\Gamma_{\rm
post}$). Figs.~\ref{fig:machcomp_circulation}(b) and~(c) demonstrate
the increasing dominance of $\Gamma_{\rm post}$ with increasing
$\chi$.

The predicted total circulation agrees better with the numerical results
when the core (rather than the cloud) value is used for $t_{\rm drag}$
in Eq.~\ref{eq:gamma_post} \citep[as also found by][]{Pittard:2009}.
Table~\ref{tab:circ} shows the predicted and measured values.

\begin{table}
\begin{center}
\caption[]{Analytical estimates for the total circulation and its
various components produced by the postshock flow.  In each case the
calculation of the value uses the drag-time for the core as given in
Table~\ref{tab:resultslo}.  $\Gamma_{\rm ring} = 0.417, 0.667,$ and
$0.743$ when $M=1.5, 3$ and 10, respectively. In the fifth column the
numerically determined peak values of the total circulation,
$-\Gamma_{\rm tot})_{\rm p}$, are given, while the sixth column notes
the dominant component(s). }
\label{tab:circ}
\begin{tabular}{llllll}
\hline
\hline
Model & $-\Gamma_{\rm shock}$ & $-\Gamma_{\rm post}$ & $-\Gamma_{\rm tot}$ & $-\Gamma_{\rm tot})_{\rm p}$ & Dom. comp.\\
\hline
m1.5c1 & 0.85 & 0.42 & 0.84 & 1.60 & $\Gamma_{\rm shock}$ and $\Gamma_{\rm post}$\\  
m3c1 & 1.37 & 0.52 & 1.22 & 1.95 & $\Gamma_{\rm shock}$ and $\Gamma_{\rm post}$ \\
m10c1 & 1.52 & 0.48 & 1.26 & 1.75 & $\Gamma_{\rm shock}$ \medskip \\

m1.5c2 & 1.12 & 6.06 & 6.76 & 6.76 & $\Gamma_{\rm post}$ \\ 
m3c2 & 1.80 & 4.20 & 5.33 & 7.48 & $\Gamma_{\rm post}$\\
m10c2 & 2.00 & 4.26 & 5.52 & 6.60$^{*}$ & $\Gamma_{\rm post}$ \medskip \\

m1.5c3 & 1.20 & 40.7 & 41.5 & 37.2 & $\Gamma_{\rm post}$ \\ 
m3c3 & 1.94 & 34.2 & 35.4 & 36.5 & $\Gamma_{\rm post}$ \\
m10c3 & 2.15 & 31.2 & 32.6 & 34.2 & $\Gamma_{\rm post}$ \\
\hline
\end{tabular}
\end{center}
\flushleft{Note: ($^{*}$) The absolute value of the total circulation
in model m10c2 was still rising at the time the simulation was
stopped.\\}
\end{table}

\subsubsection{Energy evolution}
\label{sec:global_energy}
\begin{figure}
\psfig{figure=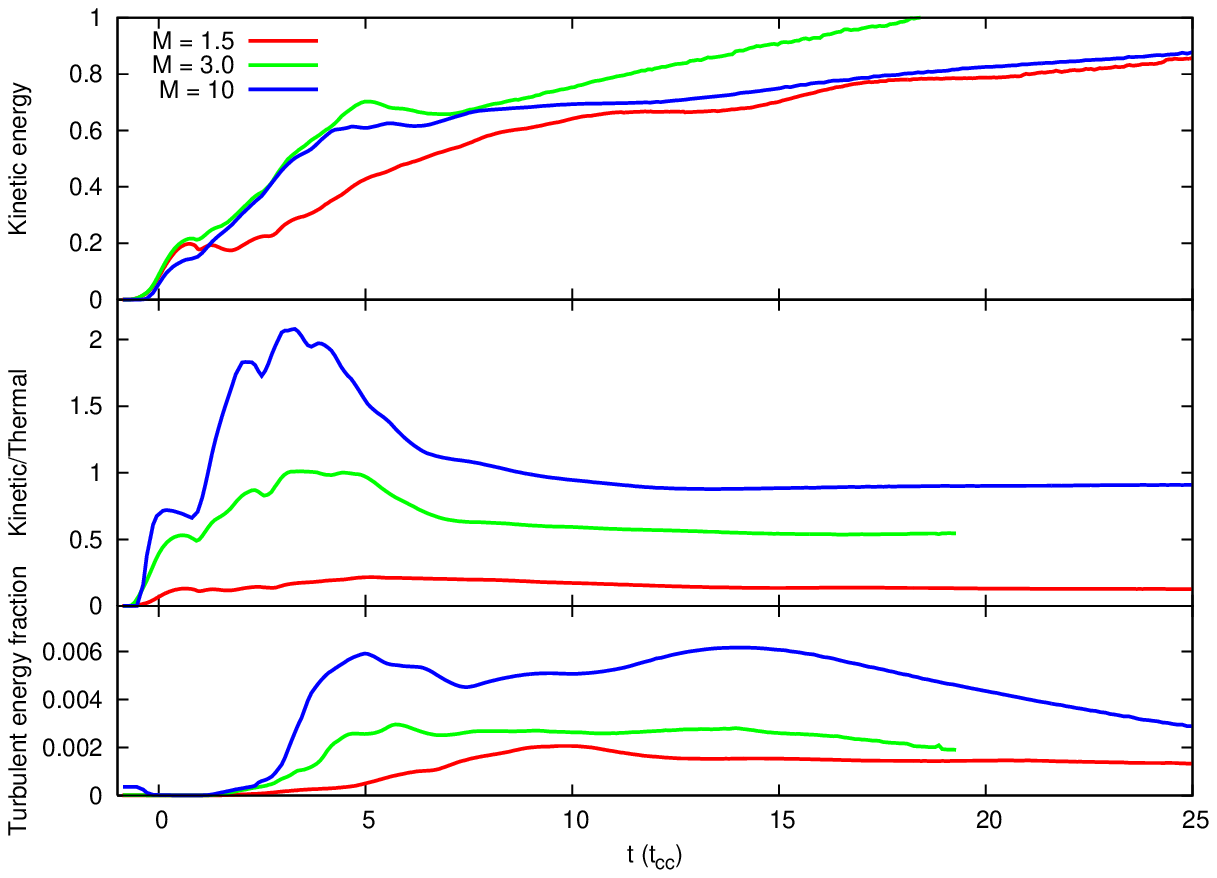,width=8.5cm}
\psfig{figure=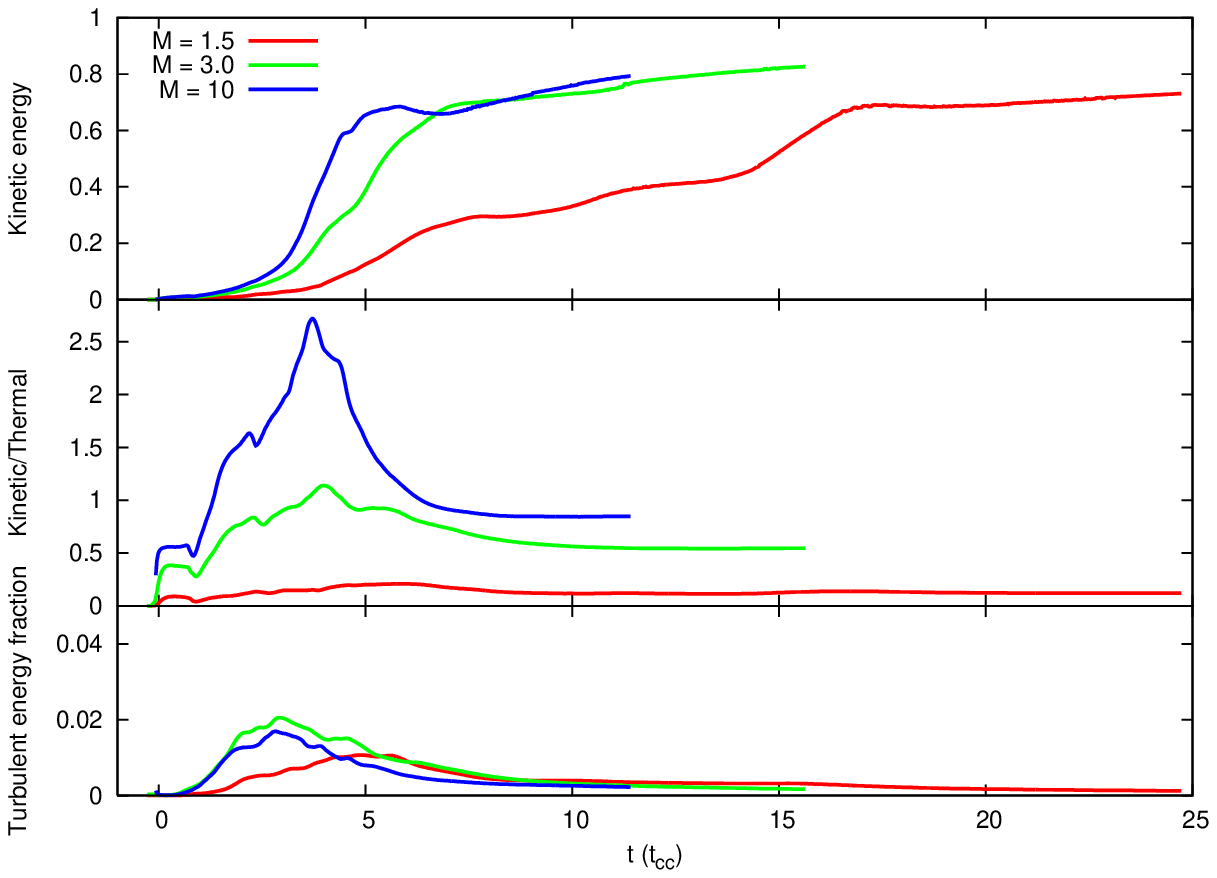,width=8.5cm}
\psfig{figure=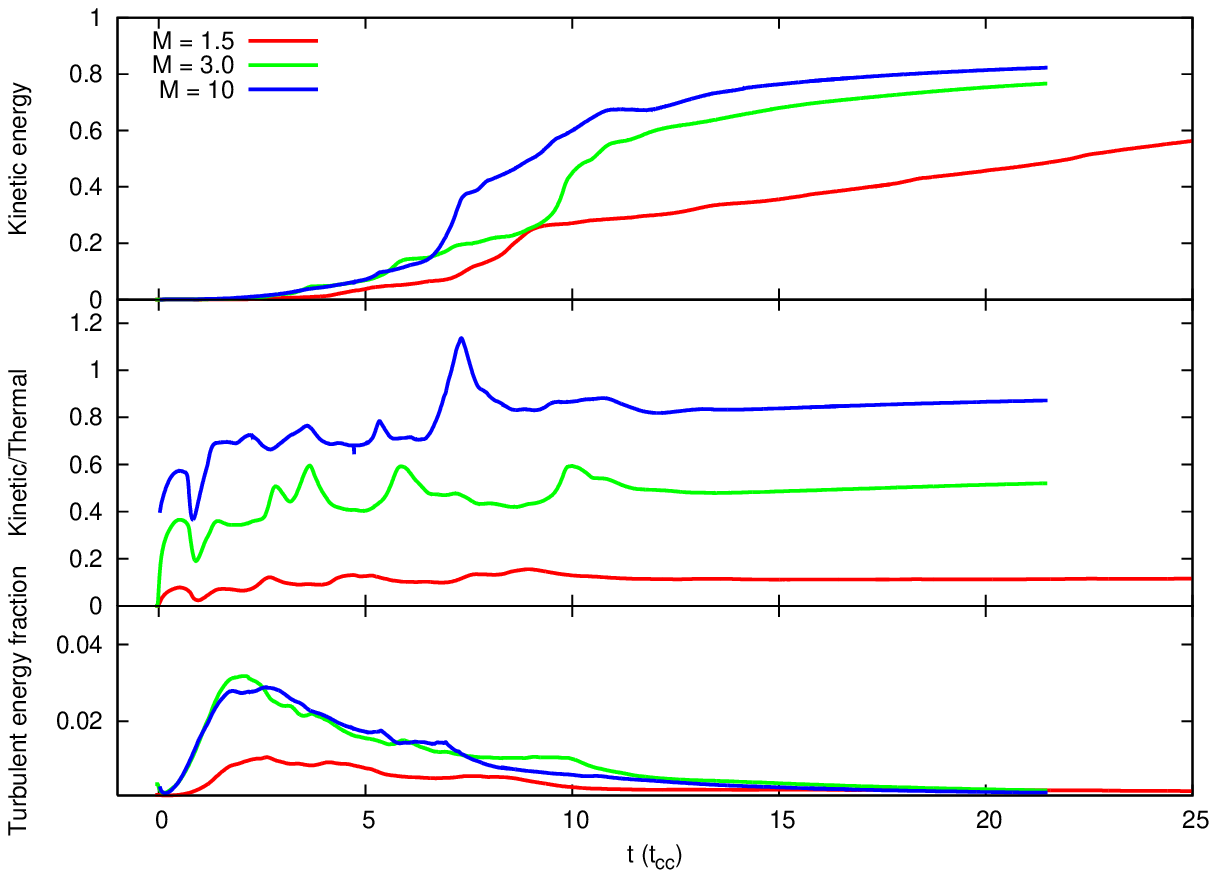,width=8.5cm}
\caption[]{The evolution of the cloud kinetic energy
(normalized to its asymptotic value once the cloud is fully mixed
into the post-shock flow) in models with
$\chi=10$ (top), $10^{2}$ (middle), and $10^{3}$ (bottom). The ratio
of the cloud's kinetic to thermal energy and its
turbulent energy at {\em subgrid scales} as a fraction of the total
cloud energy at any instant are also shown.}
\label{fig:energy_machcomp}
\end{figure}

\begin{figure}
\psfig{figure=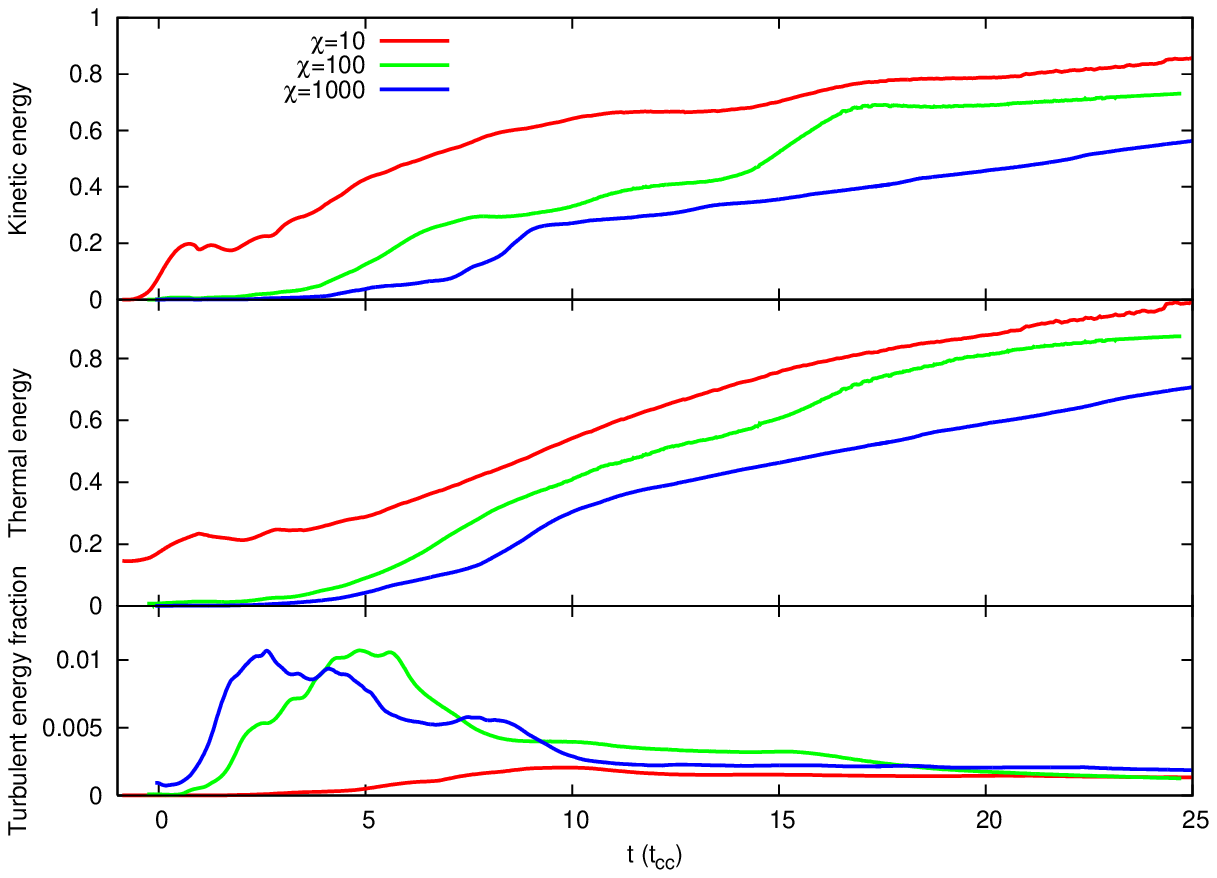,width=8.5cm}
\psfig{figure=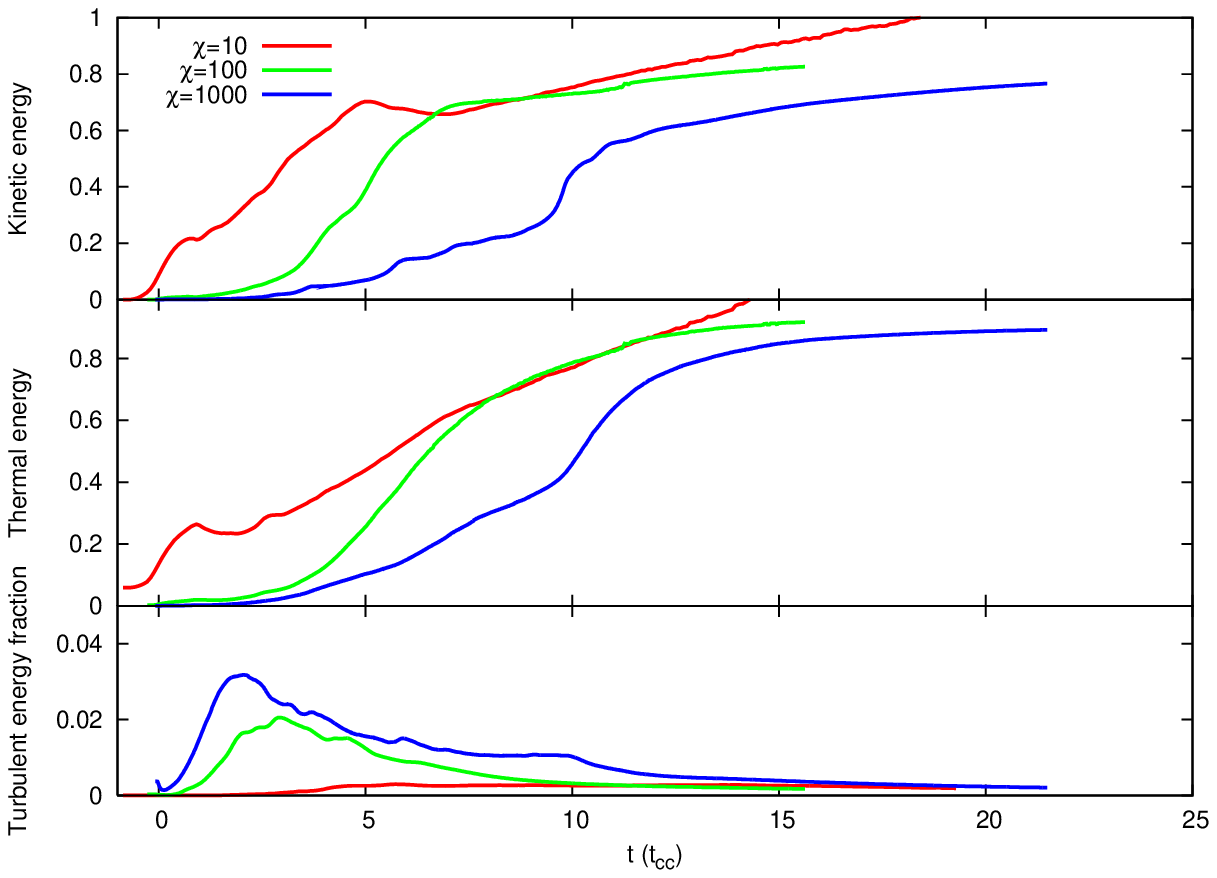,width=8.5cm}
\psfig{figure=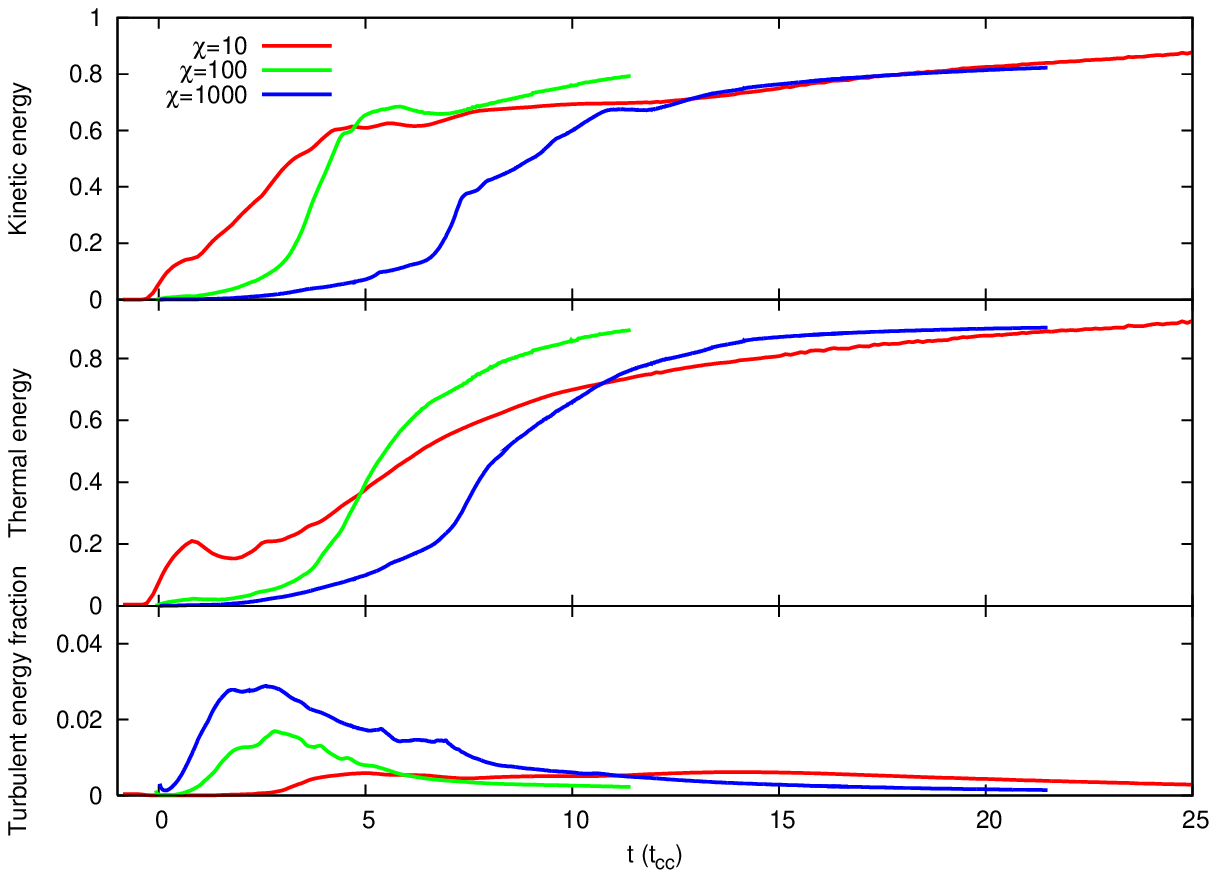,width=8.5cm}
\caption[]{As Fig.~\ref{fig:energy_machcomp} but each panel is for
models with the same shock Mach number: $M=1.5$ (top), $M=3$ (middle),
and $M=10$ (bottom). The central plot of each panel now directly
shows the thermal energy of cloud material. Note that the scale 
for the turbulent energy in the top panel differs from that in
Fig.~\ref{fig:energy_machcomp}.}
\label{fig:energy_chicomp}
\end{figure}

Figs.~\ref{fig:energy_machcomp} and~\ref{fig:energy_chicomp} show the
gain in kinetic and thermal energy of the cloud material, and the
growth (and subsequent decay) of turbulent energy for simulations as a
function of Mach number and cloud density contrast.  The cloud
material should eventually acquire the same kinetic and thermal energy
density as the ambient medium, and the turbulent energy should
dissipate as heat. The ratio of kinetic to thermal energy in the
postshock flow of an adiabatic shock is $E_{\rm k}/E_{\rm th} = \gamma
M_{\rm ps}^{2}/3$, where $M_{\rm ps}$ is the postshock Mach number
measured in the frame of the undisturbed, upstream, ambient medium. It
can be easily shown that
\begin{equation}
\label{eq:m_ps}
M_{\rm ps} = \frac{2(M^{2}-1)}{\sqrt{[2\gamma M^{2} - (\gamma - 1)][(\gamma - 1)M^{2} + 2]}}.
\end{equation}
Since $M_{\rm ps}=0.511$, 1.044 and 1.310 for $M=1.5$, $M=3$, and
$M=10$ shocks, at late times we expect that $E_{\rm k}/E_{\rm th} =
0.145$, 0.607, and 0.954 respectively. Note that $E_{\rm k}/E_{\rm th}
= 1.0$ in the strong shock (\mbox{$M \rightarrow \infty$}) limit.

The plots of the kinetic energies of the cloud material in
Fig.~\ref{fig:energy_machcomp} are normalized against the values which
the cloud material should eventually obtain once it fully mixes into,
and becomes indistinguishable from, the post-shock flow. As can be
seen, the timescales for this can be very long, especially when the
Mach number of the shock is low and the cloud density contrast is high
(when $M=1.5$ and $\chi=10^{3}$, the cloud material has attained less
than 60 per cent of its fully-mixed kinetic energy at $t=25\;t_{\rm
cc}$). 

Fig.~\ref{fig:energy_machcomp} shows that the cloud's rate of gain of
kinetic energy at a given value of $\chi$ is similar in the $M=3$ and
$M=10$ models, and in comparison is considerably slower in the $M=1.5$
models. The same behaviour is seen for the thermal energy.  In
shock-cloud interactions with $M=1.5$ the rate of energy gain is very
gradual and almost linear in time. In contrast, when $M \geq 3$, there
is rapid transfer of kinetic energy to clouds with $\chi=10$, with a
much slower transfer of thermal energy such that $E_{\rm k}/E_{\rm
th}$ can be significantly greater than unity over a substantial time
period, before finally dropping back below unity as it approaches its
asymptotic limit. The transfer of kinetic energy to clouds with
$\chi=10^{2}$ and $10^{3}$ shows a noticeable and relatively
short-lived increase in its rate when $t\sim t_{\rm drag}$ for the
cloud (this is also when $t \sim t_{\rm mix}$), after which it
slows. Fig.~\ref{fig:energy_chicomp} shows that the cloud's rate of
gain of kinetic and thermal energy at a given Mach number is initially
much more rapid for lower values of $\chi$.

The evolution of the turbulent energy is discussed in
Section~\ref{sec:turbulence}.

\subsubsection{$t_{\rm drag}$ and $t_{\rm mix}$}
\label{sec:global_tdrag_tmix}
\begin{figure}
%\begin{center}
\psfig{figure=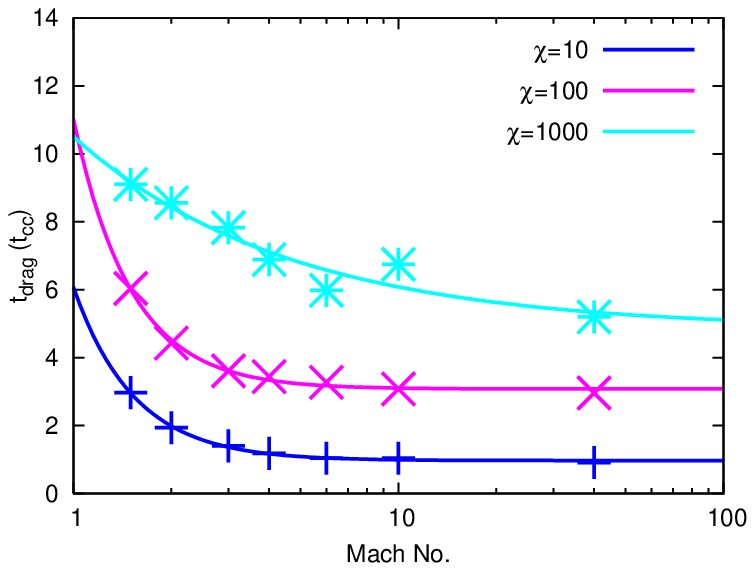,width=7.5cm}
\psfig{figure=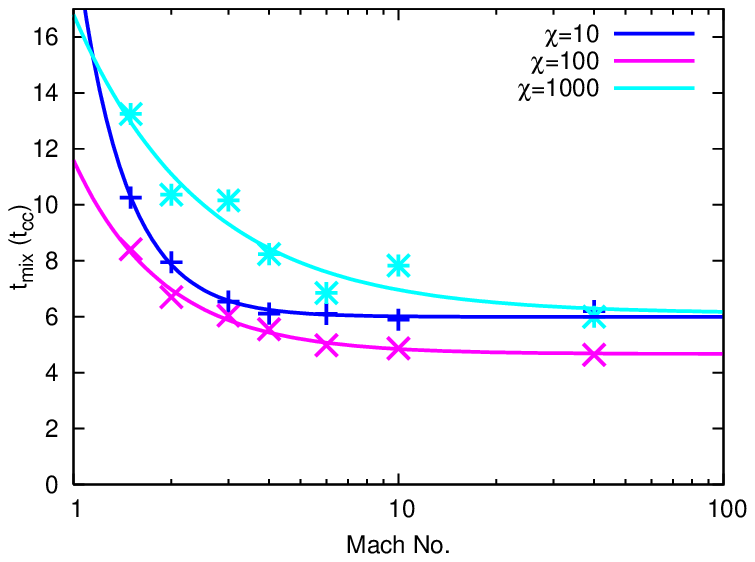,width=7.5cm}
%\end{center}
\caption[]{Top: $t_{\rm drag}$ (for the
cloud) and bottom: $t_{\rm mix}$ (for the core) as a function of Mach number 
and $\chi$ measured from the $k$-$\epsilon$ numerical simulations.}
\label{fig:tmixtdrag}
\end{figure}

Fig.~\ref{fig:tmixtdrag} shows the Mach number and $\chi$ dependence
of $t_{\rm drag}$ and $t_{\rm mix}$, two of the important timescales
which characterize the evolution and destruction of the cloud.  The
results of least-squares fits to these plots are noted in
Tables~\ref{tab:tdragfits} and~\ref{tab:tmixfits}.  Both $t_{\rm
drag}$ and $t_{\rm mix}$ are relatively constant for a given $\chi$ at
Mach numbers above 4.  This reflects the onset of the ``Mach-scaling''
mentioned in Section~\ref{sec:stages}. In contrast, $t_{\rm drag}$ and
$t_{\rm mix}$ both rise sharply at lower Mach numbers: cloud drag and
the mixing of material from the cloud are less efficient with low
shock Mach numbers. This is because: i) the density jump across
the shock and the speed of the postshock intercloud flow past the
cloud are lower when $M$ is smaller; and ii) the growth rate of
KH instabilities is slower.

The cloud is accelerated by two processes. First, the shock driven
into the cloud accelerates it to a speed $v_{\rm s}$. Further acceleration
then occurs as the shocked intercloud gas flows past the cloud, until they 
have the same velocity. For large $\chi$ the second stage dominates and
$v_{\rm s}/v_{\rm b}$ is small. Solving the equation of motion
for the cloud \citep[see, e.g., Eq.\,2.5 in][]{Klein:1994} in
the strong shock limit for $\gamma=5/3$ gives   
\begin{equation}
\label{eq:tdrag_strong}
t_{\rm drag(s)} = 1.53 \frac{\chi^{1/2}}{C_{\rm D}} t_{\rm cc},
\end{equation}
where $C_{\rm D}$ is the drag coefficient \citep[see Section~2.1
of][]{Klein:1994}. In the general case we find
\begin{equation}
\label{eq:tdrag_weak}
t_{\rm drag(g)} = 2.3 \frac{(\gamma - 1)M^{2} + 2}{(M^{2} - 1)} \frac{\chi^{1/2}}{C_{\rm D}} t_{\rm cc}.
\end{equation}
For both Eqs.~\ref{eq:tdrag_strong} and~\ref{eq:tdrag_weak}, the 
cross-sectional area of the cloud is assumed to remain constant.
This is a poor approximation which leads to an overestimate of the
drag time, since the shock causes the cloud to rapidly expand in the
transverse direction (Fig.~\ref{fig:machcomp_cloud_shape}). 
Nonetheless, it is useful to compare the ratio
$t_{\rm drag(g)}/t_{\rm drag(s)}$ with $M$. For $\gamma = 5/3$ and
$M=1.5$, we find that $t_{\rm drag(g)}/t_{\rm drag(s)} \approx 4$.
Comparing to Fig.~\ref{fig:tmixtdrag}(a) we see that the ratio of
$t_{\rm drag}$ when $M=1.5$ to when $M=40$ is 3.3, 2.0, and 1.7 for
$\chi=10$, $10^{2}$, and $10^{3}$, respectively. Hence the 
Mach number dependence of the drag time scales nearly as
expected when $\chi=10$. However, the drag time for low values of $M$
is increasingly overestimated compared to the simulations as $\chi$
increases. This divergence is at least partly due to differences
in the expansion behaviour of the cloud: comparing 
simulations m1.5c3 and m10c3, one sees that $a_{\rm cloud}$ is
significantly greater in the former until $t \approx 7\,t_{\rm cc}$
(see Fig.~\ref{fig:machcomp_cloud_shape}c).

At face value the fits in Fig.~\ref{fig:tmixtdrag}(a) to the
numerically obtained drag times imply that $C_{\rm D} \approx 5$, 5,
and 10 for $\chi=10$, $10^{2}$, and $10^{3}$ respectively.  These
values are much larger than one expects \citep[the true values of
$C_{\rm D}$ are likely to be $\ltsimm 1$ - see][for solid
bodies]{Landau:1959}.  This difference is again caused by the fact
that the cross-sectional area of the cloud does not remain constant.

Prior to the \citet{Klein:1994} paper, the naive expectation was that
the mixing time of the cloud would be comparable to the time taken for
the cloud to sweep up a column of postshock material of similar mass
to the cloud. However, \citet{Klein:1994} discovered that this was not
consistent with their results, since there was no evidence for the
mixing time scaling as $\chi^{1/2}t_{\rm cc}$, and proposed instead
that the relevant time was the timescale for KH instabilities to
fragment the cloud. The KH growth time is
\begin{equation}
t_{\rm KH} \sim \left(\frac{v_{\rm b}}{v_{\rm rel}}\right)\frac{1}{k_{\lambda}r_{\rm c}} t_{\rm cc},
\end{equation}
\noindent where $k_{\lambda}$ is the wavenumber of the perturbation.
Longer wavelengths ($k_{\lambda}r_{\rm c} \sim 1$) are the
most disruptive. Assuming that the relative speed between the
postshock flow and the cloud is the post-shock flow speed (i.e. that
the cloud is initially stationary after passage of the shock), then
$v_{\rm rel}/v_{\rm b} = (1 - \rho_{0}/\rho_{\rm s})$, where
$\rho_{0}$ and $\rho_{\rm s}$ are the preshock and postshock densities
of the ambient medium, respectively. Setting $k_{\lambda}r_{\rm c} =
1$, one then obtains
\begin{equation}
\label{eq:kh_destruction_time}
t_{\rm KH} \sim \frac{(\gamma+1)M^{2}}{(2M^{2}-2)} t_{\rm cc}.
\end{equation}
While setting $t_{\rm mix} = t_{\rm KH}$ does not give the correct
magnitude for $t_{\rm mix}$ (cf. Fig.~\ref{fig:tmixtdrag}), the
relative change with $M$ (a factor of 1.8 between $M=40$ and $M=1.5$)
is close to what is observed in our numerical model.

Fig.~\ref{fig:tmixtdrag} also shows that the mixing time of the core,
$t_{\rm mix}$, is always greater than the drag time of the cloud,
$t_{\rm drag}$. The ratio of $t_{\rm mix}/t_{\rm drag}$ is also
dependent on $\chi$, this ratio being greatest at low values of
$\chi$, but declining with increasing $\chi$ as the relative
efficiency of mixing relative to acceleration increases.
For clouds with $\chi=10$, the numerical simulations reveal that
$t_{\rm mix}/t_{\rm drag}$ increases from 1.39 at
$M=1.5$, peaking at 1.67 at $M=3$, and thereafter declines to 1.56 at
$M=40$.  When $\chi=10^{3}$, $t_{\rm mix}/t_{\rm drag}$ declines from
a value of 1.45 at $M=1.5$, to 1.15 at $M=40$. 

The values of $t_{\rm drag}$ and $t_{\rm mix}$ display greater scatter
about the least squares fit when $\chi=10^{3}$ compared to the fits at
lower values of $\chi$. This is due to large scale RT instabilites in
these models which randomly and spontaneously fragment the cloud. The
values of $t_{\rm drag}$ and $t_{\rm mix}$ are generally higher in
inviscid calculations, and display larger scatter. The extra viscosity
that the subgrid turbulence model imposes on the hydrodynamic grid
helps suppress some of the random fluctuations during the interaction,
and also leads to better convergence in resolution tests
\citep{Pittard:2009}.

\subsection{Turbulence in shock-cloud interactions}
\label{sec:turbulence}
The interaction of shocks with clouds is likely to be a key mechanism
for generating turbulence in the ISM, since substantial vorticity and
velocity dispersion is produced. Turbulent motions in our simulations
can be identified on two separate scales: the subgrid turbulence model
deals with turbulent motions and mixing on scales smaller than the
size of a grid cell, while larger scale (resolved)
turbulent motions can be directly measured from the velocity
dispersions in the axial and radial directions, $\delta v_{\rm z}$ and
$\delta v_{\rm r}$, respectively. 

\subsubsection{The cloud velocity dispersion}
Fig.~\ref{fig:machcomp_cloud_velocitydispersion} shows the time
evolution of the velocity dispersions, $\delta v_{\rm z}$ and $\delta
v_{\rm r}$.  Their behaviours are broadly similar for interactions
with supersonic post-shock flow ($M=3$ and $M=10$).  However, the
velocity dispersions in the $M=1.5$ simulations build more slowly, and
are limited to lower maximum values, reflecting the weaker
interaction.  $\delta v_{\rm z}$ peaks at higher values with $\chi$,
due to greater growth of KH and RT instabilities resulting from the
longer drag and mixing timescales with $\chi$. The ratio $\delta
v_{z}/\delta v_{\rm r}$ is not strongly dependent on $M$, in contrast
to its variation with $\chi$. The latter point was noted previously by
\citet[][]{Klein:1994} and \citet[][]{Pittard:2009}.

\subsubsection{The subgrid turbulent energy}
The time evolution of the energy in sub-grid turbulent
motions is shown in Fig.~\ref{fig:energy_machcomp}. It rises more
rapidly and peaks at higher values in simulations with high Mach
numbers and cloud density contrast.  When $\chi=10^{3}$, it peaks at
about 3 percent in models with $M=3$ and $M=10$, but peaks at only 1
percent when $M=1.5$.  In models with $\chi=10$ it peaks at less than
1 per cent, and reaches a maximum value of only 0.2 percent when
$M=1.5$. The peak in the sub-grid turbulent energy fraction is
relatively narrow when $\chi \gtsimm 10^{2}$, with most of the
turbulent energy being dissipated by $t=10\;t_{\rm cc}$. In contrast,
when $\chi=10$, the sub-grid turbulent energy can be roughly constant
for durations in excess of $15\;t_{\rm cc}$.

\subsubsection{Locally averaged velocity dispersions}
\begin{figure*}
\psfig{figure=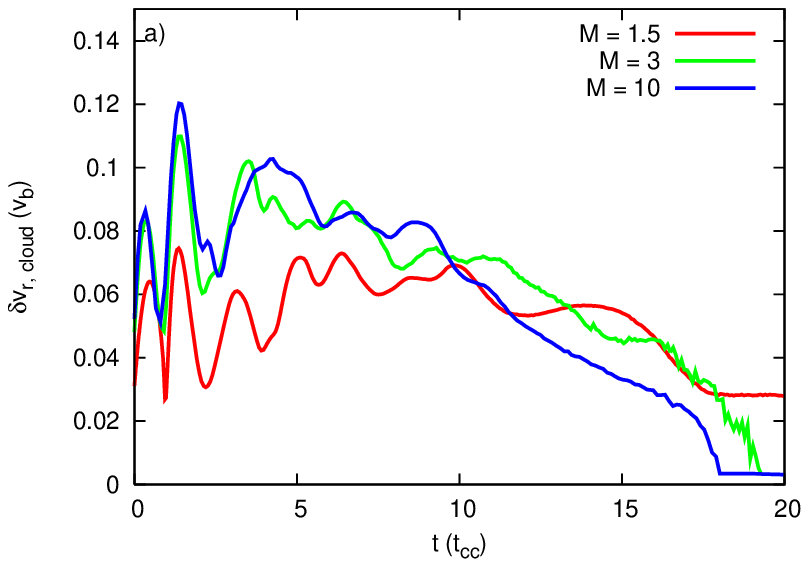,width=5.7cm}
\psfig{figure=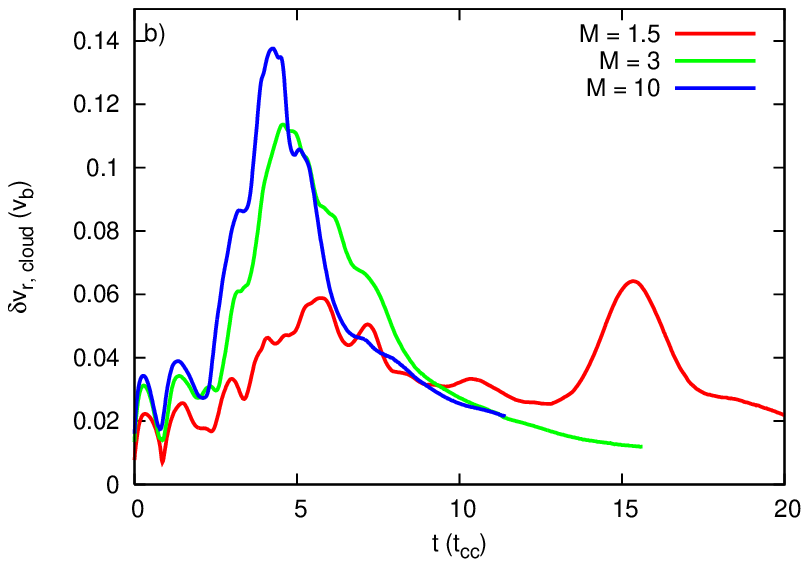,width=5.7cm}
\psfig{figure=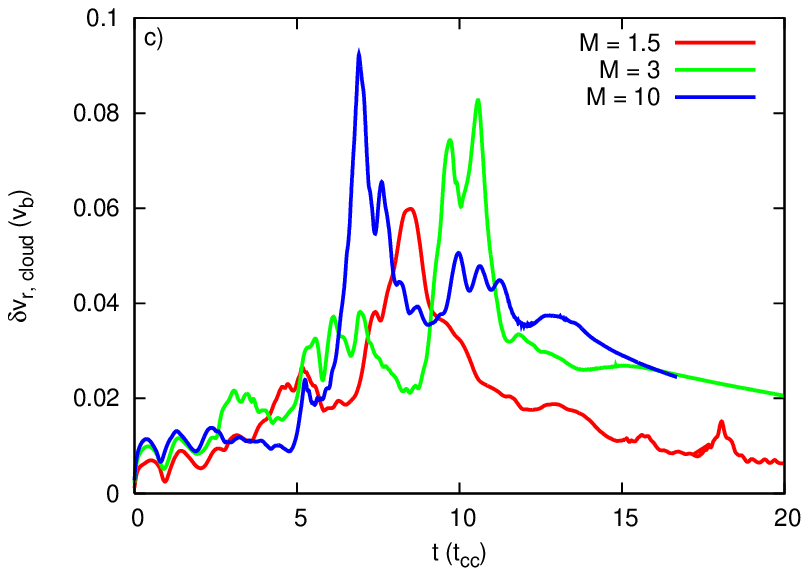,width=5.7cm}
\psfig{figure=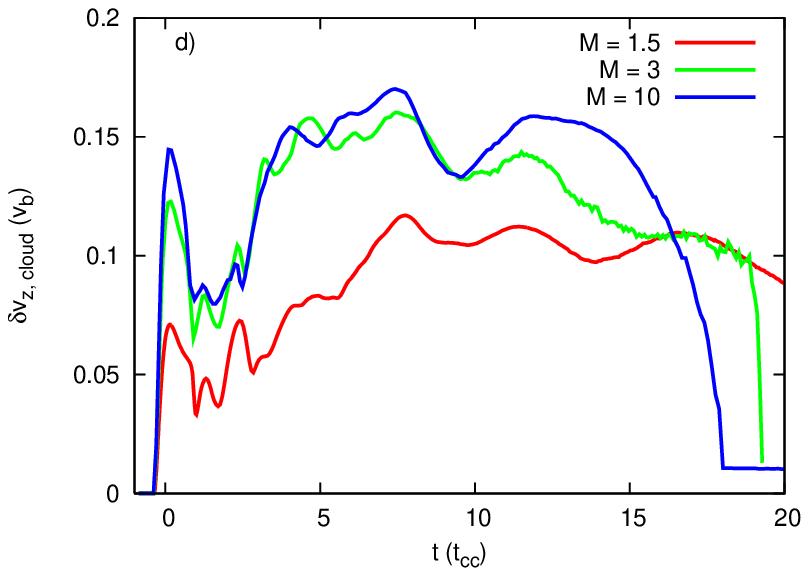,width=5.7cm}
\psfig{figure=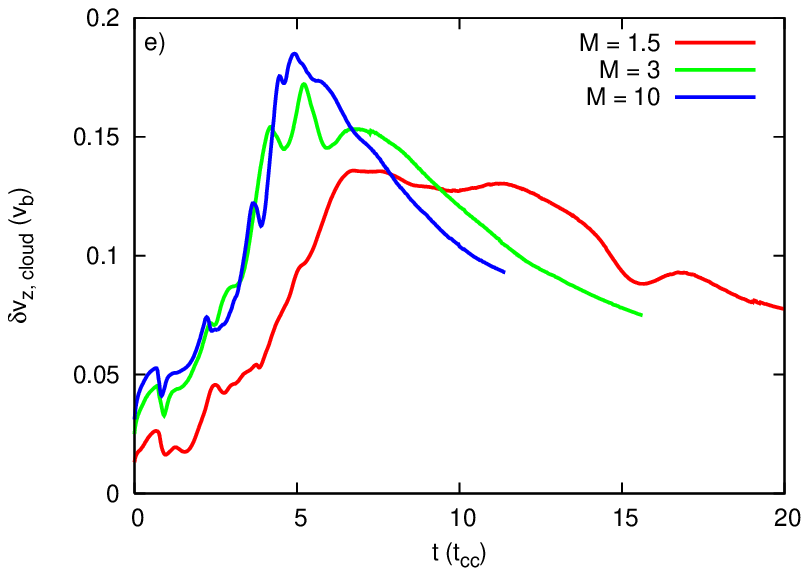,width=5.7cm}
\psfig{figure=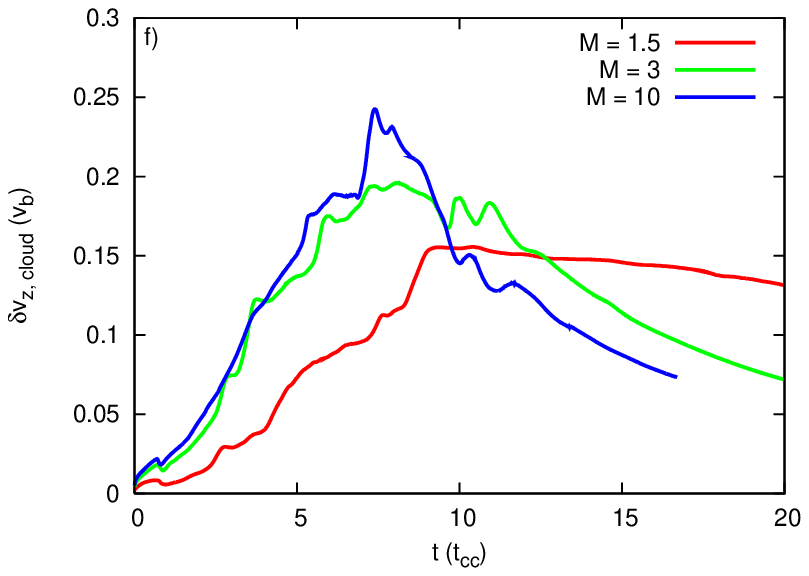,width=5.7cm}
\caption[]{Top panels: Time evolution of the cloud velocity dispersion
in the radial direction, $\delta v_{\rm r, cloud}$, for various Mach
numbers and density contrasts: (a) $\chi=10$, (b) $\chi=10^{2}$, (c)
$\chi=10^{3}$. Bottom panels: as the top panels, but for the velocity
dispersion in the axial direction, $\delta v_{\rm z, cloud}$.}
\label{fig:machcomp_cloud_velocitydispersion}
\end{figure*}

The velocity dispersions shown in
Fig.~\ref{fig:machcomp_cloud_velocitydispersion} are global averages
over the entire cloud. These can be used to obtain a globally averaged
energy fraction of the fluctuations. For model m10c3, $\delta v_{\rm
r}$ and $\delta v_{\rm z}$ peak at $t\approx 7.25\,t_{\rm cc}$, with
values of $\approx 0.09\,v_{\rm b}$ and $\approx 0.23\,v_{\rm b}$
respectively. Hence, the global velocity dispersion, $\delta v =
\sqrt{(2\delta v_{\rm r}^{2} + \delta v_{\rm z}^{2})/3} \approx
0.15\,v_{\rm b}$. This is about twice the value noted by
\citet{Nakamura:2006} for their simulations of radiative clouds with
$\gamma=1.1$. The ratio of the energy in (resolved) fluctuations to
the kinetic energy of the mean flow is $(\delta v/v)^{2} \approx 0.2$.

\begin{figure}
\psfig{figure=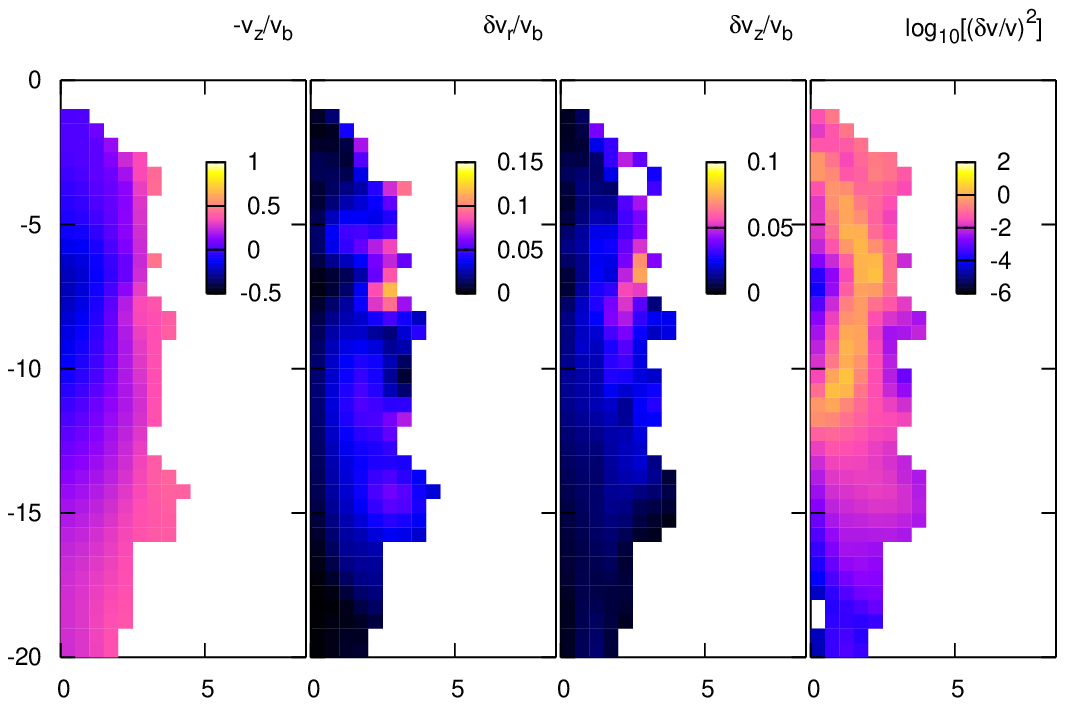,width=8.4cm}
\psfig{figure=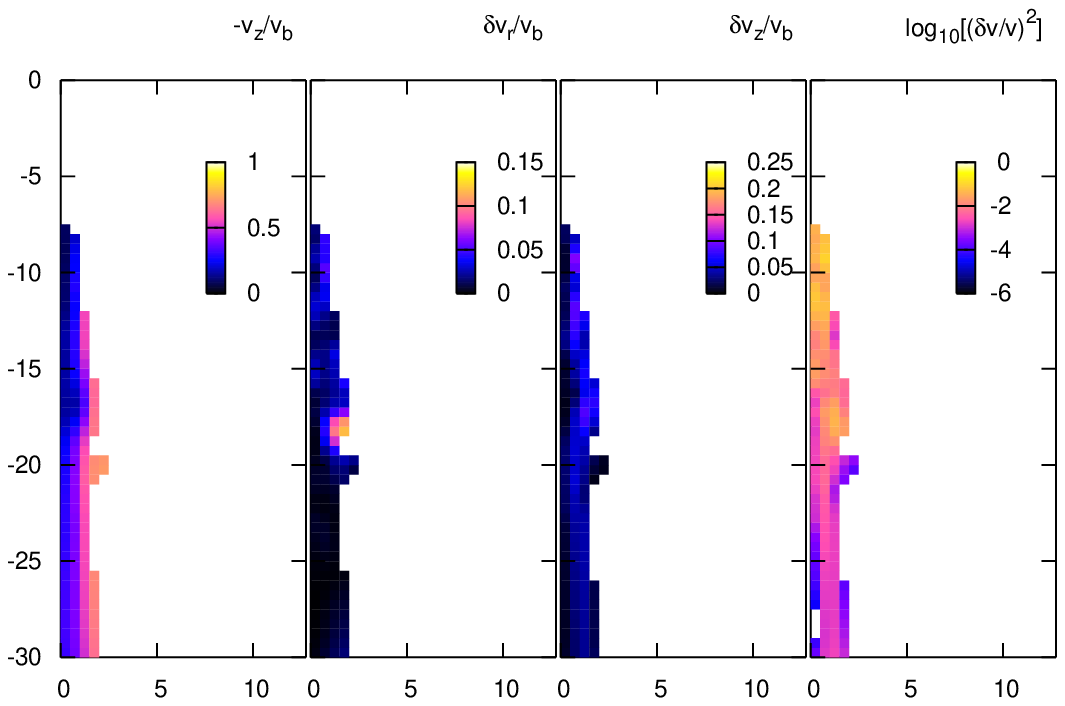,width=8.4cm}
\psfig{figure=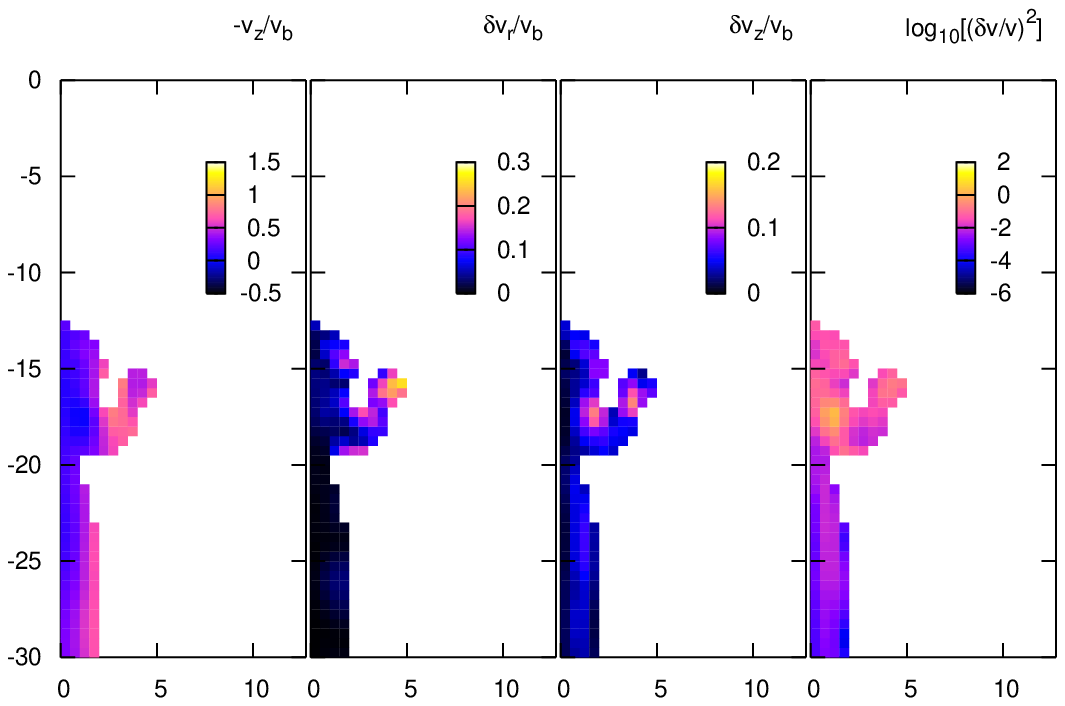,width=8.4cm}
\caption{Snapshots of $\langle v_{\rm z}\rangle$, $\delta v_{\rm r}$,
$\delta v_{\rm z}$, and ${\rm log_{10}} (\delta v/v)^{2}$. In each
pixel in these maps the mean axial velocity, the velocity dispersions,
and the ratio of the energy in (resolved) turbulent motions to the
kinetic energy of the mean flow are calculated from summing over many
smaller hydrodynamic cells. The maps were constructed at
$t=5.66\,t_{\rm cc}$ from simulations with $\chi=10^{3}$ and a)
$M=1.5$ and b) $M=10$. Panel c) shows the $M=10$ simulation at $t =
7.10\,t_{\rm cc}$, which is near the time of the peak value of the
globally averaged $\delta v_{\rm z}$ (see
Fig.~\ref{fig:machcomp_cloud_velocitydispersion}f).  The calculations
were only performed in regions containing cloud material (i.e. where the
colour in the above plots is other than white).}
\label{fig:turbulent_energy_map}
\end{figure}

However, such global estimates of $\delta v$ and $(\delta v/v)^{2}$ are
actually upper limits. Fig.~\ref{fig:turbulent_energy_map} shows the 
mean axial velocity, $v_{\rm z}$, and the velocity dispersions
$\delta v_{\rm z}$ and $\delta v_{\rm r}$, evaluated on a more
local scale. The tiles in these maps are of size 
$0.5 \times 0.5\,r_{\rm c}$, and contain up to 4096 individual
cells from the hydrodynamic calculation, over which the displayed
quantities are averaged. These averages are again
mass-weighted quantities, but consider material only within each tile 
- i.e. $\delta v_{\rm z}$ is again calculated using Eq.\,23 in
Paper~I, but $\langle v_{\rm z}^{2}\rangle$ and 
$\langle v_{\rm z}\rangle$ are evaluated only for the material
within each tile. Note that $\delta v_{\rm r}$ will be non-zero
even when $v_{\rm r}$ is constant over each tile. This is a 
consequence of the specific averaging formula. 
The maps in Fig.~\ref{fig:turbulent_energy_map}
were calculated at $t=5.66\,t_{\rm cc}$ for clouds with
$\chi=10^{3}$ hit by a Mach 1.5 (a) and a Mach 10 (b) shock (for
which the corresponding density plots are shown in Fig.~\ref{fig:machcomplo}),
and also at $t=7.10\,t_{\rm cc}$ for the $M=10$ case (panel c).

In Fig.~\ref{fig:turbulent_energy_map}(a) and (b), the mean axial
velocity is highest at the slip surface between cloud and ambient
material, and is lowest in the central core of the tail. A velocity
gradient perpendicular to the tail is clearly present. Both the $r$
and $z-$components of the velocity dispersion are also greatest at the
slip surface. Also shown is ${\rm log_{10}}(\delta v/v)^{2}$, which
has a maximum value of $-0.1$ in
Fig.~\ref{fig:turbulent_energy_map}(a), and $-0.92$ in
Fig.~\ref{fig:turbulent_energy_map}(b). Slightly later in the m10c3
simulation, a large fragment breaks away from the cloud core.  This
fragment is responsible for a significant increase in the local and
globally averaged velocity dispersions, as seen in
Figs.~\ref{fig:turbulent_energy_map}(c)
and~\ref{fig:machcomp_cloud_velocitydispersion}, respectively. The
peak local value of $(\delta v/v)^{2}=0.6$, indicating that there are
regions in the flow where the swirling motions within the gas are
almost as fast as its average bulk speed.

Our aim in this section is simply to highlight how the velocity
dispersion varies on smaller scales in the flow.  This is of interest
given that spatial variations in the velocity and the velocity
dispersion of clouds and tails can be probed using
high-spatial-resolution observations \citep[see,
e.g.,][]{Meaburn:1998,Meaburn:2009}.  We also wished to draw attention
to the fact that while the local velocity dispersion may exceed the
globally averaged value, the latter is in turn an upper limit for the
whole cloud.

\subsection{The mass-loss rate and lifetime}
\label{sec:cloud_mdot}
\begin{figure}
\psfig{figure=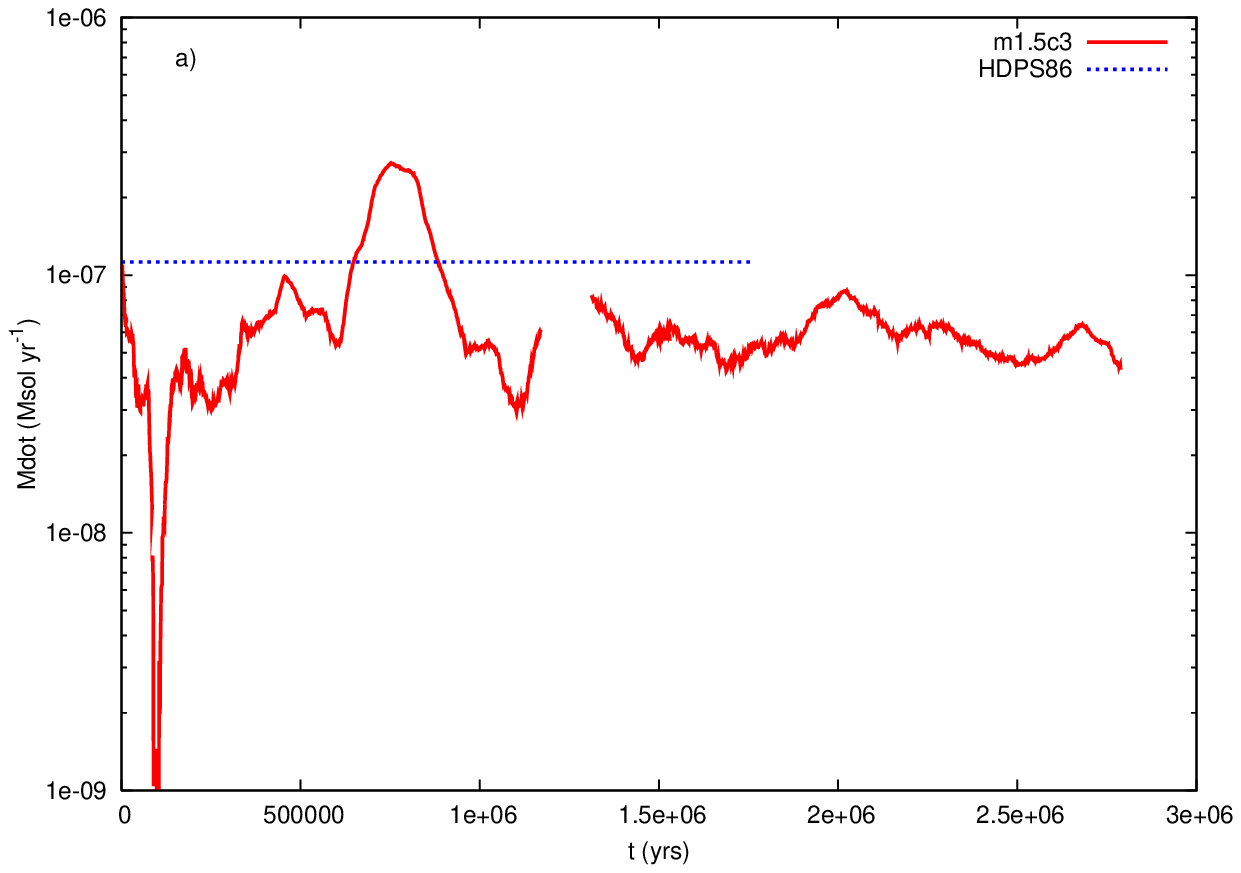,width=8.5cm}
\psfig{figure=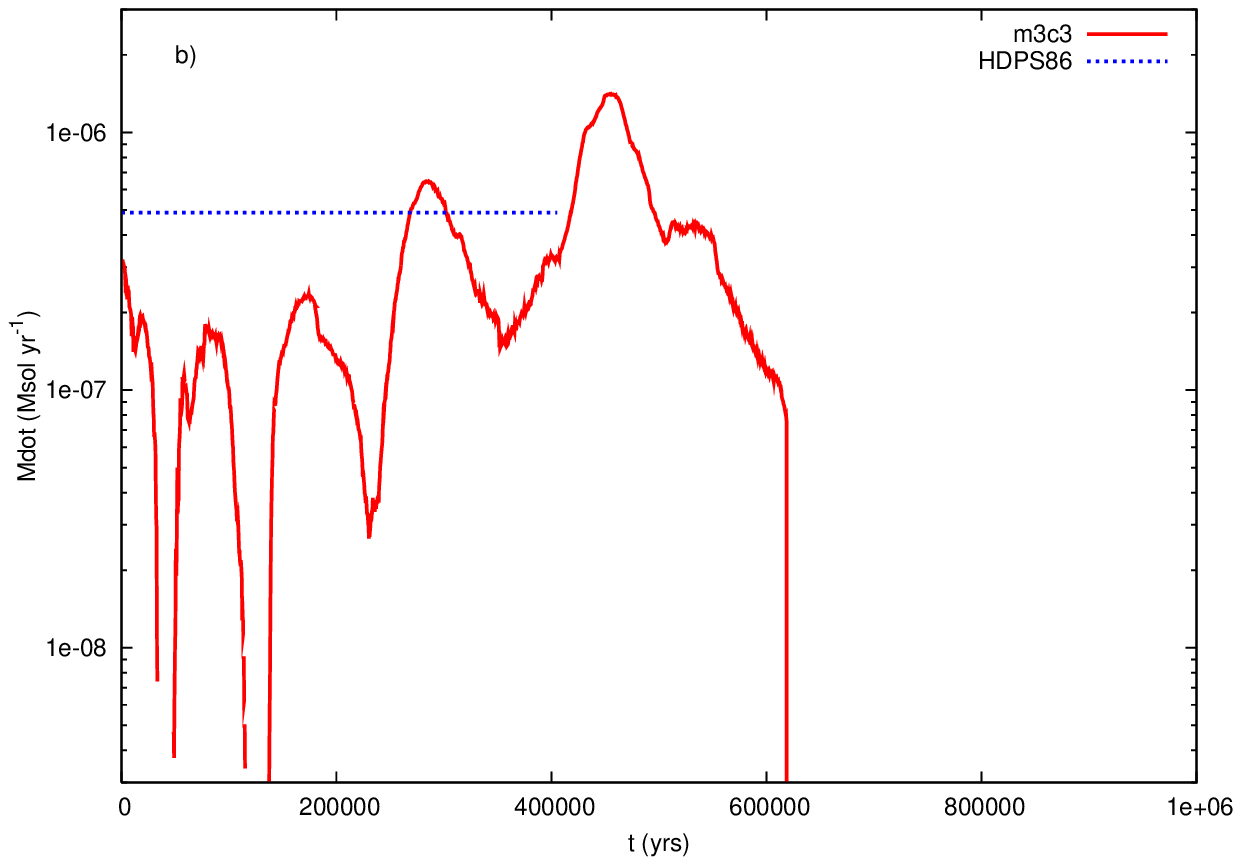,width=8.5cm}
\psfig{figure=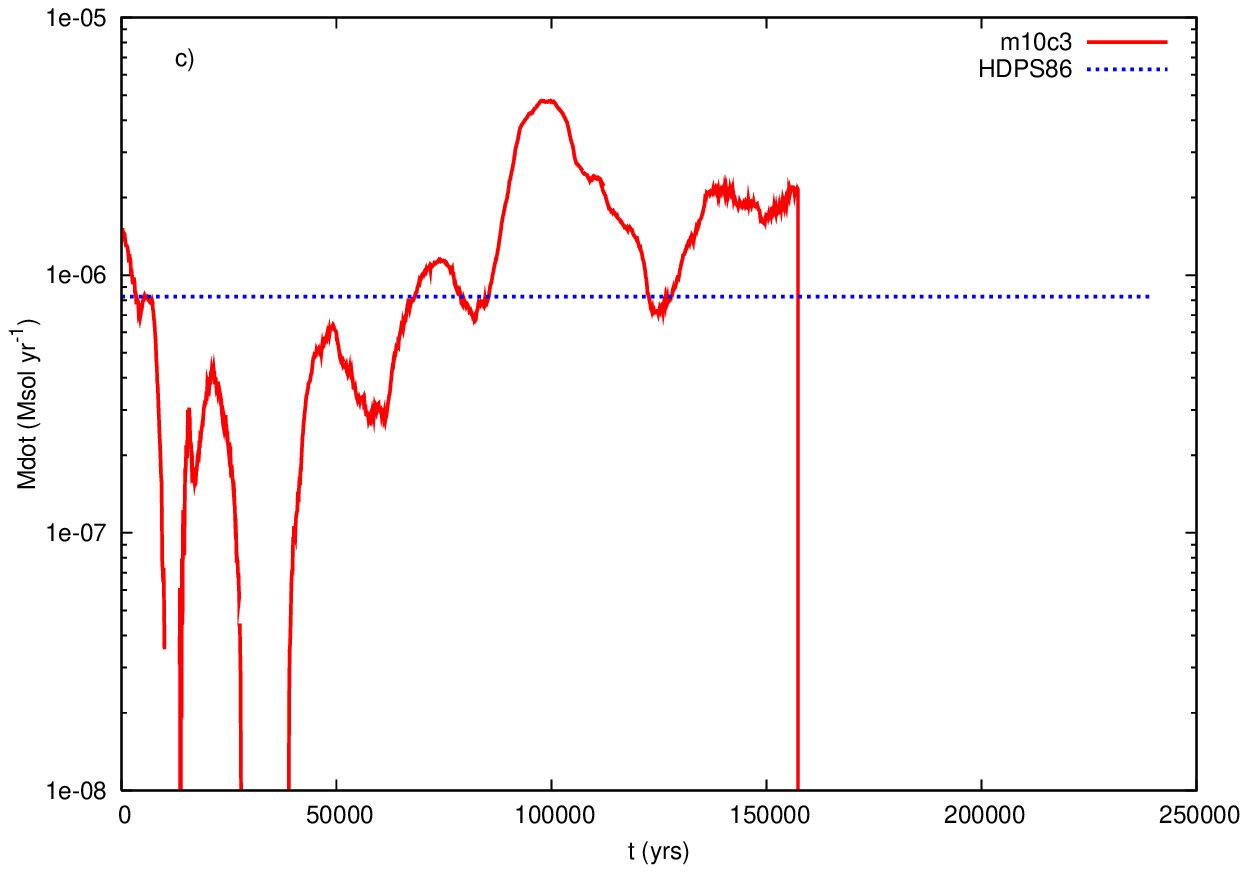,width=8.5cm}
\caption[]{Mass loss rate for models with $\chi=10^{3}$ and a)
$M=1.5$, b) $M=3$, and c) $M=10$, compared to the mass-loss formula of
\citet{Hartquist:1986}.  Gaps in the curves indicate periods where the
cloud core temporarily accretes material.}
\label{fig:massloss}
\end{figure}

The time evolution of the core mass and various mass distributions have 
been previously discussed in Section~\ref{sec:cloud_mass}. 
Here we note a more detailed examination of this mass-loss.
In Fig.~\ref{fig:massloss} the rate of mass-loss from the core is
compared to the analytical formula for hydrodynamic ablation given by
\citet{Hartquist:1986}. For this comparison we assume that the
cloud is fully ionized with a radius $r_{\rm c} = 2\;$pc, core density
$\rho_{\rm c}=4\times10^{-25}\;{\rm g\;cm^{-3}}$, and temperature of
$8000\;$K, and is in pressure equilibrium with the surrounding gas of density
$4\times10^{-28}\;{\rm g\;cm^{-3}}$ (i.e. $\chi=10^{3}$) and
temperature $8 \times 10^{6}\;$K. Both the cloud and its surroundings
are assumed to have solar abundances (average mass per particle,
$\mu=0.61\,m_{\rm H}$).  A shock of Mach number 1.5, 3, or 10, then
travels through the ambient medium at a speed of 650, 1300, or
$4300\;\kmps$, heating the medium to 12, 29, or 260 million~K.

The analytical formula proposed by \citet{Hartquist:1986} depends on
the Mach number of the flow past the cloud. If this flow is supersonic
(such as when the shock Mach number, $M > 2.76$), then $\Mdot_{\rm ab}
\approx (M_{\rm c} c_{\rm c})^{2/3} (\rho v)^{1/3}$, while if the Mach
number of the flow past the cloud, $M_{\rm ps}$, is subsonic, then
there is an additional $M_{\rm ps}^{4/3}$ dependence.  $M_{\rm c}$ is
the mass of the cloud, $c_{\rm c}$ is the sound speed within it, and
$\rho$ and $v$ are respectively the density and velocity of the
environment external to the cloud (the post-shock ambient/intercloud
gas).  With the above parameters, $\Mdot_{\rm ab} \approx 1.1$, 4.9,
and $8.2 \times 10^{-7}\;\Msolpyr$ for interactions with Mach number
1.5, 3, and 10 shocks, and the cloud survives for approximately 1.8,
0.41, and 0.24 million yrs. In comparison the cloud crushing
timescale, $t_{\rm cc} = 9.5$, 4.8, and $1.4 \times 10^{4}\;$yrs, so
that the cloud survives for about 18 cloud crushing timescales before
being destroyed (clouds with a lower density contrast survive for a
smaller multiple of $t_{\rm cc}$).

From Fig.~\ref{fig:massloss} we see that the level of agreement
between $\Mdot_{\rm ab}$ and the numerically determined mass-loss
rates from the simulations is reasonably good (gaps in the numerical
curves indicate short periods when the cloud core accretes
material).  Furthermore, the rate of mass-loss from the core shows
less extreme variations as the Mach number is reduced (i.e. the cloud
is destroyed in a gentler fashion).  However, the agreement shown in
Fig.~\ref{fig:massloss} is actually rather fortuitous, as we now
demonstrate.  The lifetime of the cloud, $t_{\rm life} = M_{\rm
c}/\Mdot_{\rm ab}$, can be expressed in units of the cloud crushing
timescale, $t_{\rm cc}$, as
\begin{eqnarray}
\label{eq:tlife}
t_{\rm life}/t_{\rm cc} & = & \left(\frac{M_{\rm c} c_{\rm c}}{\rho v}\right)^{1/3} \frac{M c_{\rm amb}}{r_{\rm c} c_{\rm c}^{2/3} \chi^{1/2} [M_{\rm ps}^{4/3}]}\nonumber \\ 
 & = & 1.6 \left(\frac{\rho_{\rm c} c_{\rm c}}{\rho v}\right)^{1/3} \frac{M}{[M_{\rm ps}^{4/3}]},
\end{eqnarray}
where the term in square brackets should only be used if the
post-shock flow of density $\rho$ and velocity $v$ over the cloud is
subsonic. As \mbox{$M \rightarrow \infty$},
because of the linear dependence of $v$ on $M$, one finds that $t_{\rm
life}/t_{\rm cc} \propto M^{2/3}$.  For constant $\rho_{\rm amb}$ and
$c_{\rm amb}$ (as in the simulations), Eq.~\ref{eq:tlife} gives
$t_{\rm life}/t_{\rm cc} \propto (\rho_{\rm c} c_{\rm c})^{1/3}$,
and, since $\rho_{\rm c} = \chi \rho_{\rm amb}$ and 
$c_{\rm c} = \chi^{-1/2} c_{\rm amb}$, we find that $t_{\rm
life}/t_{\rm cc} \propto \chi^{1/6}$.

\begin{figure}
\begin{center}
\psfig{figure=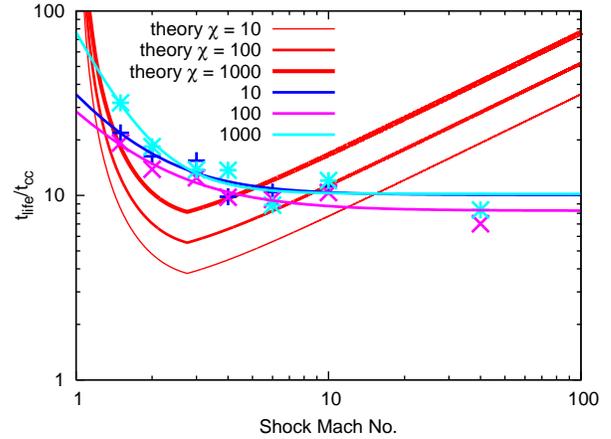,width=8.5cm}
\end{center}
\caption[]{Comparison of the cloud lifetime from the numerical
simulations (defined as the time when the ``core'' disappears)
and the lifetime obtained from the analytical result of
\citet{Hartquist:1986}, as a function of the shock Mach number, $M$,
and the cloud density contrast, $\chi$.}
\label{fig:tlife}
\end{figure}

The ratio of the core lifetime from the numerical
models\footnote{Defined as the time at which the core disappears -
i.e. when material originally in the core has at least as much ambient
material in each computational cell on the hydrodynamic grid.} to the
lifetime assuming steady mass-loss at the rate using the analytical
formula of \citet{Hartquist:1986} is shown in Fig.~\ref{fig:tlife}.
The kink seen in the analytical curves at $M=2.76$ reflects the switch
from Mach number dependent mass-loss when the postshock flow is
subsonic ($M < 2.76$ - i.e. the inclusion of the $M_{\rm ps}^{4/3}$
term in the above equations) to Mach number independent mass-loss when
the postshock flow is supersonic. The displacement of the
$\chi=10^{2}$ and $\chi=10^{3}$ curves reflects the $\chi^{1/6}$
proportionality noted above.

The agreement for clouds with density contrasts $\chi\sim10^{3}$ hit
by shocks with Mach numbers $M\ltsimm 10$ is again good, as was shown
previously in Fig.~\ref{fig:massloss}. However, there is a significant
and increasing divergence between the numerical and analytical cloud
lifetimes as the shock Mach number increases past $M=15$. For a Mach
40 shock, the numerical results indicate that the cloud is
``destroyed'' by $t\approx 8\,t_{\rm cc}$, whereas the analytical
formula suggests a cloud lifetime of $20-40\,t_{\rm cc}$ for
$\chi=10-10^{3}$, or up to 5 times longer. This is because of the
$M^{2/3}$ scaling of Eq.~\ref{eq:tlife} at high Mach numbers.  Note
that the numerical results are consistent with Mach scaling
(Section~\ref{sec:stages}), whereas the analytical formula is
not. Perhaps even more serious is the disagreement at moderate Mach
numbers ($M\ltsimm7$) and low density contrasts ($\chi\sim10$).  In
such cases the cloud survives appreciably longer than the analytical
formula of \citet{Hartquist:1986} suggests. For $\chi=10$ and $M=3$,
the analytical formula suggests that the cloud will survive until
$t=4\,t_{\rm cc}$, whereas the numerical calculation suggests the
cloud actually survives until about $t=15\,t_{\rm cc}$ (i.e.  its
mass-loss rate is about 4 times lower than the equation in Hartquist
et al. would suggest).

These differences have consequences for previous mass-loading
calculations in which the analytical mass-loss rate prescription from
\citet{Hartquist:1986} was adopted \citep*[see,
e.g.,][]{Dyson:1987,Arthur:1996a,Strickland:2000,Dyson:2002}.
However, in practice, we believe that the results from such
calculations will actually change very little, for a number of
reasons. Firstly, mass-loading causes a flow's Mach number to tend
towards unity or thereabouts: supersonic flows are slowed as a result
of momentum transfer and their temperature and sound speed increased
through frictional heating, while subsonic flows are accelerated by
mass-loading (see Eq.~5 in \citet{Hartquist:1986}, and the numerical
calculations of \citet*{Arthur:1993}). This means that although use of
the formula in \citet{Hartquist:1986} for a cloud overrun by a high
Mach number shock or immersed in a high Mach number flow results in
the initial mass-loading being too slow, the rapid decrease in the
Mach number of that flow in response to this mass-loading means that
the analytically determined mass injection rate from clouds which are
further downstream is closer to the correct (numerically determined)
value. This behaviour is seen, for instance, in the mass-loading of
the high Mach number pre-termination-shock stellar wind material in
wind-blown-bubbles \citep*{Pittard:2001a}. It is also seen in the
simulations of \citet{Arthur:1993} and \citet*{Arthur:1996b}, where
the position of the reverse shock of the wind-blown-bubble is fixed at
the radius of the onset of mass-loading. Secondly, in most of these
works values for the clouds (such as $\chi$, $r_{\rm c}$, etc.) were
not explicitly specified. Instead, the authors simply adopt a constant
rate of mass-injection into supersonic flow, modified by an $M^{4/3}$
dependence for subsonic flow. Such simulations are therefore
``exempt'' from the large disparity beteen the analytical and
numerical results for $\chi\ltsimm 10$ and $M \ltsimm 7$ which we have
discovered in this work\footnote{Obviously other works examining
different types of mass-loading in diffuse sources, for example
through thermal conduction
\citep*[e.g.][]{McKee:1977,Chieze:1981,White:1991,Pittard:2001b,Pittard:2003a,Pittard:2003b,Pittard:2004}
or photoevaporation
\citep*[e.g.][]{Garcia-Arredondo:2002,Pittard:2005}, or where the
mass-loading rate is unrelated to the properties of the local flow
\citep*[e.g.][]{Smith:1996,Dyson:1995,Williams:1995,Williams:1999}, or
is not from embedded clouds
\citep*[e.g.][]{Weaver:1977,Toniazzo:2001}, remain unaffected.}.

Having established the existence of a large difference between the
analytical and numerical lifetimes of clouds at high Mach numbers and
at low-to-moderate Mach numbers for clouds with small density
contrasts, we now wish to know the underlying cause of this
discrepancy.  For a flow streaming {\em supersonically}
(i.e. \mbox{$M_{\rm ps} > 1$}) past a cloud, \citet{Hartquist:1986}
assumed that mass-loss occurs largely as a result of low pressure
regions on the surface of the cloud. The mass-loss rate and lifetime
of the cloud (in cgs units) is then found to be Mach number
independent. However, this leads to a Mach number dependent lifetime
for the cloud when expressed in units of $t_{\rm cc}$
(Eq.~\ref{eq:tlife}), since the dependence is introduced by $t_{\rm
cc}$. In contrast, the numerical simulations presented in this work
(and earlier simulations in the literature) show that, for clouds with
$\chi \gtsimm 10^{2}$, material is initially stripped from the sides
of the cloud through the development of KH instabilities, while the
core remains relatively intact. At later times, and at early times for
clouds with $\chi \ltsimm 10$, the instabilities break up the cloud
into several large fragments. It seems reasonable to conclude,
therefore, that the mass-loss process identified by
\citet{Hartquist:1986} is in fact not the dominant mode of mass-loss
from deformable clouds subject to large velocity shear.

In contrast, Eq.~\ref{eq:kh_destruction_time} shows that the
destruction time due to long wavelength KH instabilities has only a
weak dependence on the shock Mach number. One finds that $t_{\rm KH} =
2.4 \,t_{\rm cc}$ when $M=1.5$, and $t_{\rm KH} = 1.33 \,t_{\rm cc}$
when $M=40$. These estimates are roughly 6 times shorter than the
numerically determined lifetimes at high shock Mach number ($M\gtsimm
7$). The scaling of the cloud lifetime (in units of $t_{\rm cc}$) with
$M$ is also weaker than is observed from the simulations (see
Fig.~\ref{fig:massloss}). For instance, it predicts that the lifetime
of a cloud hit by a Mach 1.5 shock is about 1.8 times longer than in a
Mach 40 interaction.  Nonetheless, the insensitivity of
Eq.~\ref{eq:kh_destruction_time} to the cloud density contrast,
$\chi$, is consistent with the numerical results, where only a weak
(and non-montonic) dependence is seen (Fig.~\ref{fig:massloss}). Given
these observations, it seems reasonable to associate the mode of cloud
destruction with KH instabilities, but it is clear that the lifetime
of the cloud, $t_{\rm life} \sim 6\,t_{\rm KH}$. This expression is
within a factor of 2 or so of the numerically determined lifetime of
clouds with $10 < \chi < 10^{3}$ hit by shocks with $M > 1.5$, with
the biggest discrepancy at low Mach numbers. The parameters of the
fits to the numerical results are noted in Table~\ref{tab:tlifefits}.

\begin{table}
\begin{center}
\caption[]{The parameters from least squares fits to $t_{\rm
drag}/t_{\rm cc}$ for the cloud (see Fig.~\ref{fig:tmixtdrag}a). The
final column lists the sum of the residuals. In each case the fitted
function has the form $t_{\rm drag}/t_{\rm cc} = aM^{b} + c$.  Each fit
has 4 degrees of freedom.}
\label{tab:tdragfits}
\begin{tabular}{lcccc}
\hline
\hline
$\chi$ & $a$ & $b$ & $c$ & $\Sigma\;{\rm res}^{2}$ \\
\hline
10 & $5.11\pm0.35$ & $-2.33\pm0.16$ & $0.97\pm0.03$ & 0.0487 \\
$10^{2}$ & $7.94\pm0.84$ & $-2.46\pm0.25$ & $3.08\pm0.07$ & 0.043 \\
$10^{3}$ & $5.67\pm0.86$ & $-0.66\pm0.34$ & $4.84\pm1.04$ & 0.956 \\
\hline
\end{tabular}
\end{center}
\end{table}

\begin{table}
\begin{center}
\caption[]{The parameters from least squares fits to $t_{\rm mix}/t_{\rm
cc}$ for the cloud (see Fig.~\ref{fig:tmixtdrag}b). 
The final column lists the sum of the
residuals. In each case the fitted function has the form $t_{\rm
mix}/t_{\rm cc} = aM^{b} + c$.  Each fit has 4 degrees of freedom.}
\label{tab:tmixfits}
\begin{tabular}{lcccc}
\hline
\hline
$\chi$ & $a$ & $b$ & $c$ & $\Sigma\;{\rm res}^{2}$ \\
\hline
10 & $13.9\pm1.67$ & $-2.90\pm0.28$ & $6.00\pm0.09$ & 0.086 \\
$10^{2}$ & $6.93\pm0.69$ & $-1.59\pm0.23$ & $4.67\pm0.15$ & 0.135 \\
$10^{3}$ & $10.7\pm1.95$ & $-1.10\pm0.43$ & $6.11\pm0.96$ & 2.81 \\
\hline
\end{tabular}
\end{center}
\end{table}

\begin{table}
\begin{center}
\caption[]{The parameters from least squares fits to $t_{\rm life}/t_{\rm
cc}$ for the cloud. The final column lists the sum of the
residuals. In each case the fitted function has the form $t_{\rm
life}/t_{\rm cc} = aM^{b} + c$.  Each fit has 4 degrees of freedom.}
\label{tab:tlifefits}
\begin{tabular}{lcccc}
\hline
\hline
$\chi$ & $a$ & $b$ & $c$ & $\Sigma\;{\rm res}^{2}$ \\
\hline
10 & $25.2\pm10.1$ & $-1.89\pm1.06$ & $10.1\pm2.00$ & 13.9\\
$10^{2}$ & $20.3\pm4.94$ & $-1.62\pm0.56$ & $8.26\pm1.06$ & 6.52\\
$10^{3}$ & $66.6\pm21.6$ & $-2.82\pm0.76$ & $10.2\pm1.20$ & 16.3 \\
\hline
\end{tabular}
\end{center}
\end{table}

\section{Summary and conclusions}
\label{sec:summary}
This is the second of a series of papers investigating the turbulent
destruction of clouds. In our first paper \citep{Pittard:2009} the
focus was primarily on the much faster evolution and dispersal of a
cloud over-run by a shock with a highly turbulent post-shock flow.
Here we have performed a detailed examination of the Mach number
dependence of the destruction of a cloud by an adiabatic shock.  We
have used a hydrodynamical code which incorporates a sub-grid
compressible $k$-$\epsilon$ turbulence model in an attempt to
calculate the properties of the turbulence and the resulting increase
in the transport coefficients.

We find that the most significant differences in the nature of the
interaction occur between those where the ambient post-shock flow is
subsonic with respect to the cloud, and those where it is supersonic
(the latter requires that the shock Mach number $M > 2.76$). For this
reason, the interaction of a Mach 3 shock is more akin to a Mach 10
shock than a Mach 1.5 shock. At high Mach numbers the post-shock
conditions are virtually independent of the Mach number and the
so-called ``Mach scaling'' is obtained. Mach scaling appears to hold
also when using the $k$-$\epsilon$ turbulence model.

For weak shocks ($M < 2.76$) the interaction is much milder and the
following differences are observed: i) the postshock flow is subsonic
with respect to the cloud, so a bowwave rather than a bowshock forms
ahead of the cloud, ii) the compression of the cloud is more
isotropic, iii) a weaker vortex ring is produced, iv) the smaller
velocity difference at the slip surface around the cloud limits the KH
and RT instabilities and reduces the peak turbulent energy fraction of
the flow, v) it takes much longer for the cloud to be mixed into the
surrounding flow and for it to accelerate to the intercloud postshock
speed, and vi) mass stripped from the cloud does not as readily form a
long tail (the set-up time is longer). We further find that a
prominent tail only forms if $\chi \gtsimm 10^{3}$.

Our most important finding is that the analytical prescription in
\citet{Hartquist:1986} for the ablative mass-loss rate of a cloud in
an external flow predicts cloud lifetimes which are in disagreement
with numerically determined values.  For instance, the predicted
lifetimes are a factor of $2-5$ times too long for clouds with $10 <
\chi < 10^{3}$ hit by a Mach 40 shock, while they are about 4 times
too short for clouds with $\chi=10$ hit by an $M=3$ shock. The reason
for these discrepancies appears to be due to the assumption in
\citet{Hartquist:1986} that the mass-loss is mostly driven by pressure
gradients around the cloud. Instead, we show that the cloud lifetime
is more closely related to the timescale for large scale KH
instabilities, though it is about 6 times longer than the latter.  We
argue, however, that the rapid reduction in the Mach number of
hypersonic flows subject to mass-loading means that previous work in
the literature is unlikely to greatly change if repeated using a more
accurate mass-loss rate prescription.

In future work we will extend our investigation to three dimensions,
examine the interaction of a dense shell with a cloud, and will
compare synthetic signatures of the interaction to the types of
diffuse sources mentioned in the introduction.

\section*{acknowledgements}
We would like to thank the referee for a helpful report which improved
this paper. JMP would also like to thank the Royal Society for funding
a University Research Fellowship, and is grateful for useful discussions 
with John Dyson on some of this work.

\label{lastpage}

\end{document}